\documentclass{article}

\usepackage{arxiv}

\usepackage[utf8]{inputenc} 
\usepackage[T1]{fontenc}    
\usepackage{hyperref}       
\usepackage{url}            
\usepackage{booktabs}       
\usepackage{amsfonts}       
\usepackage{nicefrac}       
\usepackage{microtype}      
\usepackage{lipsum}		
\usepackage{graphicx}
\usepackage{natbib}
\usepackage{doi}
\usepackage{amsthm}
\usepackage{amsmath}
\usepackage{color}
\usepackage{xcolor}
\usepackage{colortbl}
\definecolor{C1}{HTML}{93BFCF}
\definecolor{C2}{HTML}{A0C3D2}
\definecolor{C3}{HTML}{BDCDD6}
\definecolor{C4}{HTML}{EEE9DA}
\definecolor{C5}{HTML}{FFF1DC}
\definecolor{C6}{HTML}{E8D5C4}
\definecolor{C7}{HTML}{EEEEEE}
\definecolor{C8}{HTML}{BCEE68}
\usepackage{multirow}
\usepackage{hhline}
\usepackage{bbding}

\title{Towards Unified Representation of Multi-Modal Pre-training for 3D Understanding via Differentiable Rendering}

\date{} 					

\author{Ben Fei$^{*,1}$, Yixuan Li$^{*,1}$, Weidong Yang{$^{\dagger,1}$}, Lipeng Ma, Ying He{$^{\dagger, 2}$}\\
	$^1$Fudan University, $^2$Nanyang Technological University\\
    \texttt{bfei21@m.fudan.edu.cn}, \texttt{wdyang@fudan.edu.cn},
    \texttt{yhe@ntu.edu.sg}\\
}

\renewcommand{\shorttitle}{\textit{} DR-Point}

\begin{document}
\maketitle

\begin{abstract}
	State-of-the-art 3D models, which excel in recognition tasks, typically depend on large-scale datasets and well-defined category sets. Recent advances in multi-modal pre-training have demonstrated potential in learning 3D representations by aligning features from 3D shapes with their 2D RGB or depth counterparts. However, these existing frameworks often rely solely on either RGB or depth images, limiting their effectiveness in harnessing a comprehensive range of multi-modal data for 3D applications. To tackle this challenge, we present \textbf{DR-Point}, a tri-modal pre-training framework that learns a unified representation of RGB images, depth images, and 3D point clouds by pre-training with object triplets garnered from each modality. To address the scarcity of such triplets, DR-Point employs differentiable rendering to obtain various depth images. This approach not only augments the supply of depth images but also enhances the accuracy of reconstructed point clouds, thereby promoting the representative learning of the Transformer backbone. Subsequently, using a limited number of synthetically generated triplets, DR-Point effectively learns a 3D representation space that aligns seamlessly with the RGB-Depth image space. Our extensive experiments demonstrate that DR-Point outperforms existing self-supervised learning methods in a wide range of downstream tasks, including 3D object classification, part segmentation, point cloud completion, semantic segmentation, and detection. Additionally, our ablation studies validate the effectiveness of DR-Point in enhancing point cloud understanding.
\end{abstract}

\keywords{Self-supervised learning, contrastive learning, point cloud understanding, multi-modal, differentiable rendering.}

\newcommand\blfootnote[1]{%
\begingroup
\renewcommand\thefootnote{}\footnote{#1}%
\addtocounter{footnote}{-1}%
\endgroup
}
\blfootnote{{$*$}Equal contribution, {$\dagger$}Corresponding author.}

\section{Introduction}
\label{sec:intro}

\begin{figure}[ht]
    \centering
    \includegraphics[width=\linewidth]{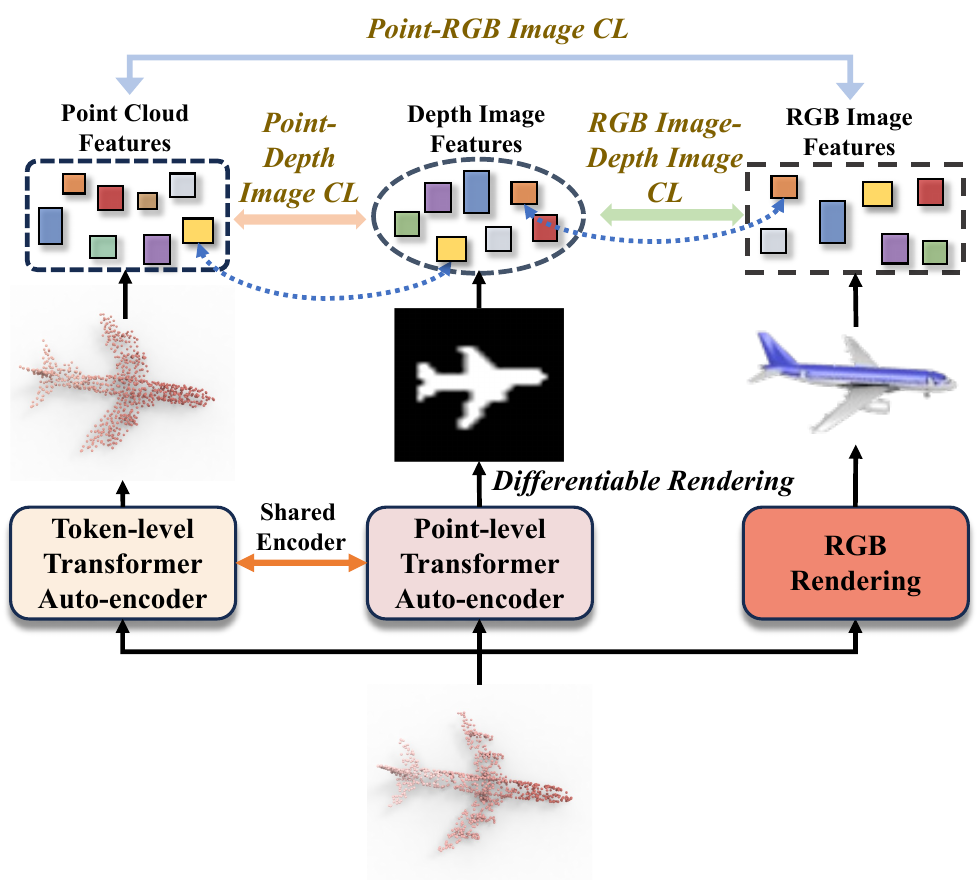}
    \vspace{-0.9cm}
    \caption{Illustrations of DR-Point, a methodology for improving the 3D understanding by aligning features from tri-modalities, such as RGB images, depth images, and point clouds into a shared space. DR-Point aims to reduce the requirement of object triplets using \textbf{Differentiable Rendering} to obtain depth images, together with RGB images and point clouds from image-3D pairs to enhance the representative learning of models. 
    }
    \label{fig:teaser}
\end{figure}

Due to the ever-growing demand for real-world applications in augmented/virtual reality, autonomous driving, and robotics, 3D visual understanding has attracted increasing attention in recent years~\citep{fei2023self}.
However, compared to their 2D counterpart, 3D visual recognition remains restricted due to the presence of small-scale datasets and a limited range of pre-defined categories~\citep{zhang2022point}.
The limited scalability of 3D data presents a significant obstacle to the widespread adoption of 3D recognition models and their practical applications. This limitation arises due to the substantial expenses associated with both the collection and annotation of 3D data~\citep{zhang2022pointclip}.

In addressing the scarcity of annotated data, previous research in various domains has demonstrated that leveraging knowledge from diverse modalities can greatly enhance the understanding of the original modality. 
Among these, CrossPoint~\citep{afham2022crosspoint} pioneered alignment between 2D and 3D features via a cross-modal contrastive learning approach to learn transferable 3D point cloud representations. 
It demonstrates superior performance compared to previous unsupervised learning methods across a wide range of downstream tasks, such as 3D object classification and segmentation.
Otherwise, CLIP2Point~\citep{huang2022clip2point} integrates cross-modality learning to leverage depth features for capturing both visual and textual expressions, as well as intra-modality learning to enhance the invariance of depth aggregation.
However, these methods contrast RGB images or depth images from different views, determining that they are relatively easy to be aligned.
Moreover, learning from either RGB images or depth images will make the models concentrate on the texture information from RGB images or edge information from depth images.

On the other hand, many generative methods in self-supervised learning endeavor to recover point clouds from masked ones.
As a pioneering work, Point-BERT~\citep{yu2022point} implements mask language modeling, inspired by BERT, in the context of 3D data. It utilizes a dVAE to tokenize 3D patches, randomly masks certain 3D tokens, and then predicts them during the pre-training phase.
Taking a step further, PointMAE~\citep{pang2022masked} directly operates on point cloud by masking out 3D patches and predicting the masked patches.
Although they are effective in downstream tasks, uni-modal reconstruction loss like Cross-Entropy (CE) or Chamfer Distance (CD) is inadequate for capturing various geometric details in original data.

In this paper, to tackle the above-mentioned challenges, we propose learning a unified representation of RGB images, depth images, and point clouds (\textbf{DR-Point}). 
An illustration of our framework is shown in Fig.~\ref{fig:teaser}, where three branches are carefully devised.
(i) Following MaskPoint~\citep{liu2022masked}, a Token-level Transformer Auto-encoder (TTA) is integrated to reconstruct the masked point clouds at token-level and point features can be obtained in this branch;
(ii) Further, to enhance the representative learning ability of the shared Transformer encoder, Point-level Transformer Auto-encoder (PTA) is devised to recover the masked point clouds at point-level via CD loss.
Moreover, we design a differentiable rendering to promote the accuracy of reconstructed point clouds.
It can be regarded as a ``kill two birds with one stone'' method, which obtains depth image features by a feature extractor.
(iii) In the last branch, we embed the corresponding rendered 2D image into feature space. 
After acquiring tri-modal features, the point cloud, RGB images, and depth images can be embedded closely to one another within the feature space. This embedding ensures the preservation of the correspondence between the Point-RGB-Depth (PRD).

The joint tri-modal learning objective compels the model to achieve several desirable attributes. 
Firstly, it enables the model to identify and understand the compositional patterns present in three modalities. 
Secondly, it allows the model to acquire knowledge about the spatial and semantic properties of point clouds by enforcing invariance to modalities.
After undergoing tri-modal pre-training without any manual annotation, the pre-trained encoder can be effectively transferred to various downstream tasks. Our DR-Point showcases superior performance, as demonstrated through a comprehensive comparison against widely recognized benchmarks.

\section{Related Works}
\label{sec:related}

\begin{figure*}[ht]
    \centering
    \includegraphics[width=\linewidth]{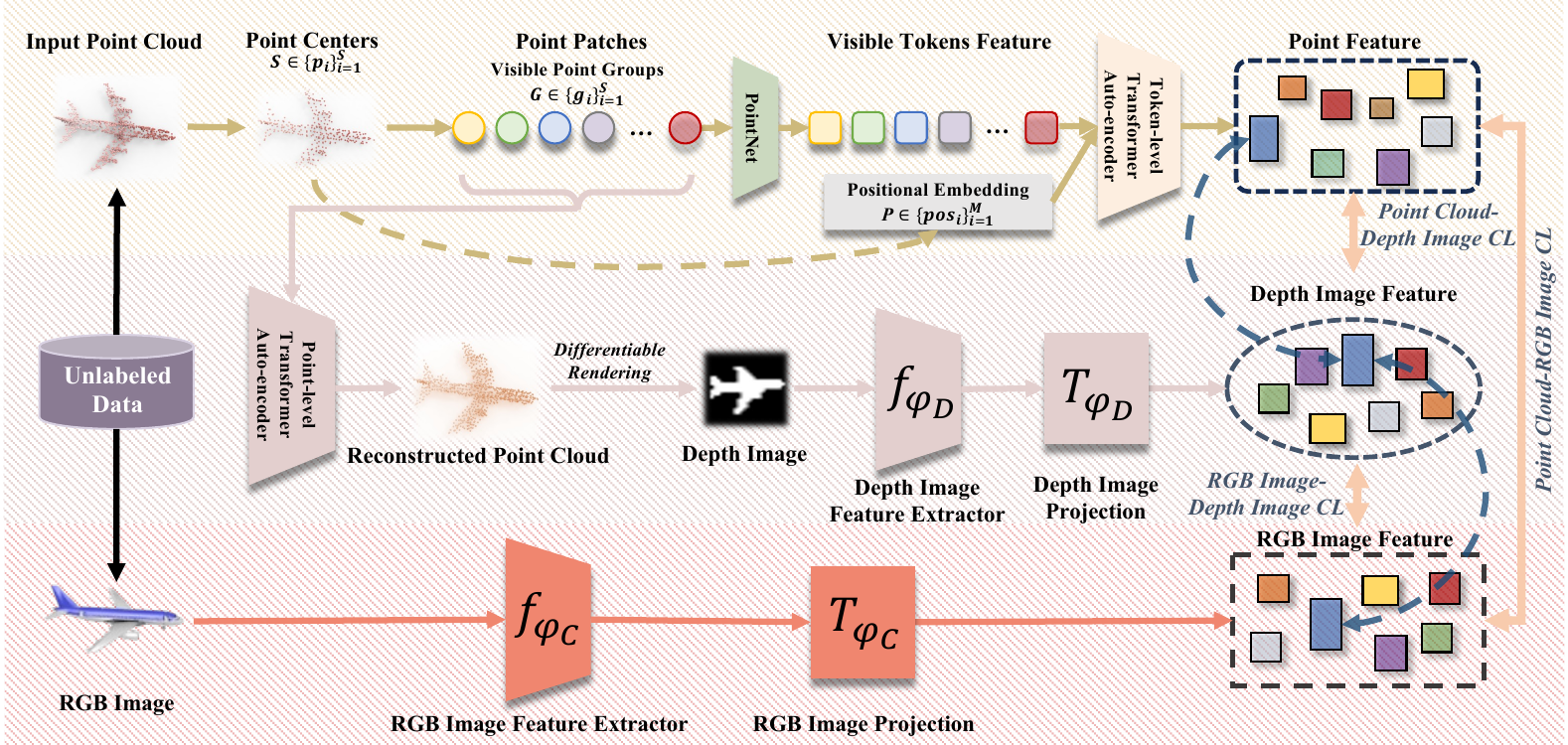}
    \vspace{-0.7cm}
    \caption{Illustration of DR-Point. 
    The tri-modal pre-training of DR-Point requires a batch of objects represented as triplets (RGB image, depth image, point cloud), which are extracted from three branches:
    (i) Token-level Transformer Auto-encoder (\textbf{Top}) aims to recover point clouds at the token level as well as exploit 3D features;
    (ii) Point-level Transformer Auto-encoder (\textbf{Middle}) is designed to reconstruct point clouds at the point-level, which shares the Transformer encoder with the former branch.
    Moreover, differentiable rendering is leveraged to ensure the reconstruction of high-quality point clouds from 32 random views, while one random depth view will be leveraged to exploit depth features;
    (iii) RGB features (\textbf{Bottom}) are extracted from a pre-trained ResNet with a projection head. 
    During pre-training, contrastive losses are applied to align the 3D feature of an object with its corresponding RGB and depth features. }
    \label{fig:overview}
\end{figure*}

\textbf{Multi-model Pre-training.}
Most existing multi-modal approaches leverage image and text modalities into point cloud understanding~\citep{mao2023complete,yan2023comprehensive}. 
One particular set of methods, including CLIP, employs image and text encoders to generate a unified representation for each image-text pair. These representations from both modalities are subsequently aligned.
The simplicity of this architecture enables efficient training with large amounts of noisy data, thereby facilitating its ability to generalize even in zero-shot scenarios.
The success of CLIP has led to a proliferation of research related to the integration of images and text~\citep{wang2023nerf,zheng2024learning}.
Some recent works explore how multi-modal information can help 3D understanding and show promising results. 
For instance, PointCLIP~\citep{zhang2022pointclip} first transforms the 3D point cloud into a collection of depth maps. Subsequently, it directly utilizes CLIP for zero-shot 3D classification.
The other researches focus on aligning image modalities. 
CrossPoint~\citep{afham2022crosspoint} aims at establishing a 3D-2D correspondence of objects by optimizing the alignment between point clouds and their respective rendered 2D images within the invariant space, while CLIP2Point~\citep{huang2022clip2point} integrates cross-modality learning to enhance the depth features, enabling the capture of rich visual and textual characteristics and intra-modality learning is employed to improve the invariance of depth aggregation.
In contrast to the approaches presented in CrossPoint~\citep{afham2022crosspoint} and CLIP2Point~\citep{huang2022clip2point}, our proposed method, DR-Point, enables the acquisition of a comprehensive and integrated representation across RGB images, depth images, and point clouds, resulting in significant advancements in 3D comprehension.

\noindent\textbf{3D Point Cloud Understanding.} 
There are primarily two research directions focused on the understanding of point clouds~\citep{zhang2023pointvst}. 
On the one hand, supervised learning tends to project a point cloud into 3D voxels and and subsequently utilizes 2D/3D convolutions to extract features~\citep{wang2019dynamic}. 
Further, PointNet~\citep{qi2017pointnet} and PointNet++~\citep{qi2017pointnet++} explore processing 3D point clouds directly. 
The PointNet architecture effectively captures permutation-invariant features from point clouds, which have a substantial impact on point-based 3D networks. In contrast, PointNet++~\citep{qi2017pointnet++} introduces a hierarchical neural network that progressively extracts local features with varying contextual scales.
Recently, PointMLP~\citep{ma2022rethinking} proposes a pure residual MLP network that achieves competitive results without the need for complex local geometric extractors.
On the other hand, self-supervised learning has demonstrated promising performance in the field of 3D understanding for point clouds~\citep{fei2023self}. 
PointBERT~\citep{yu2022point} applies the concept of mask language modeling from BERT to the domain of 3D understanding. In this approach, 3D patches are tokenized using an external model, and random tokens are masked. The model is then trained to predict the masked tokens during the pre-training phase.
Built on top of Point-BERT, PointMAE~\citep{pang2022masked}, focuses on direct manipulation of point clouds. PointMAE involves masking 3D patches within the point cloud and predicting their 3D positions using the CD loss.
However, the unified model reconstruction loss, such as the Cross-Entropy loss in Point-BERT or the Chamfer Distance loss in Point-MAE, is insufficient for capturing the diverse geometric intricacies present in the original 3D data.
Our DR-Point aims to solve this challenge by devising tri-modal pre-training to learn a more universal representation.

\noindent\textbf{Differentiable Rendering.} 
Differentiable rendering techniques are widely employed in 3D reconstruction tasks, allowing for the generation of rendering images. These techniques also support 3D model reconstruction through back-propagation.
There are four categories of existing differentiable renderers based on geometric representation: point-based~\citep{roveri2018pointpronets,grigoryan2004point}, voxel-based~\citep{lin2018learning,gan2023v4d}, mesh-based~\citep{hermosilla2018monte,correa2009comparison}, and implicit neural function-based~\citep{sitzmann2019scene,chubarau2023cone} approaches.
Voxel-based methods~\citep{lin2018learning} necessitate substantial memory allocation for lower-resolution geometries, whereas mesh-based methods~\citep{hermosilla2018monte} leverage the sparsity of 3D geometry.
However, converting geometries into meshes is challenging and prone to errors. These methods have limitations in terms of global and topological alterations, and their connectivity lacks differentiability. 
Implicit neural functions have gained popularity as a means of representing high-resolution scenes. However, existing approaches~\citep{sitzmann2019scene} encounter limitations in terms of network capacity and the accurate alignment of camera rays with scene geometry.
Point-based methods~\citep{roveri2018pointpronets} operate directly on point samples of the geometry, making it both a flexible and efficient approach. 
Hence, the integration of a proficient point-based differentiable renderer enables the capture of rendering images from diverse camera angles, thereby facilitating local geometry reconstruction and our tri-modal pre-training.

\section{Method}
\label{sec:DR-Point}
DR-Point (Fig.~\ref{fig:overview}) is pre-trained on triplets extracted from RGB images, depth images, and 3D point clouds, learning a unified representation space of these different modalities.
This section will introduce the creation of triplets for pre-training (Sec.~\ref{sec:Triplets}) as well as our pre-training framework (Sec.~\ref{sec:Align}).

\subsection{Creating Training Triplets for DR-Point}
\label{sec:Triplets}
As the availability of training triplets is often limited, it becomes necessary to generate them during the pre-training process.
On one hand, the acquisition of rendered RGB images is facilitated by the synthetic nature of the pre-training dataset. However, it should be noted that these RGB images are directly rendered from 3D objects, making them easily alignable.
On the other hand, the depth images are always unavailable in the pre-training dataset. Consequently, it is imperative to develop a real-time depth renderer in order to acquire the depth images on the fly.

\noindent\subsubsection{RGB Image Rendering} 
The rendered RGB images are sourced from~\citep{xu2019disn}, which comprises 43,783 images depicting 13 distinct object categories.
A 2D image is randomly chosen from all rendered images for each point cloud, captured at an arbitrary viewpoint.
Each point cloud consists of 2,048 points and a corresponding rendered RGB image resized to 224 $\times$ 224.
To enhance the complexity of the pre-training task and enhance the models' meaningful representations, it is essential to subject the rendered RGB images to data augmentation. This will make the alignment of the tri-modal data more challenging.
Data augmentation for rendered images includes random crop, color jittering, and random horizontal flips.
After undergoing data augmentation, RGB images are utilized as input in the RGB branch (Shown in Fig.~\ref{fig:overview} (\textbf{Bottom})).

\noindent\subsubsection{Depth Image Generation via Differentiable Rendering}
To tackle the unavailable depth images in the pre-training dataset, drawing inspiration from Insafutdinov et al.~\citep{insafutdinov2018unsupervised}, a differentiable rendering loss is devised.
The incorporation of differentiable rendering is implemented within the point-level transformer auto-encoder branch. This branch seeks to reconstruct the 3D positions of point clouds (Shown in Fig.~\ref{fig:overview} (\textbf{Middle})), which serve as the input for the differentiable renderer.
Hence, differentiable rendering not only enhances the reconstruction of point-level transformer auto-encoder from their respective projections using depth image modality, but also efficiently generates depth images during the pre-training process. 

\begin{figure}[t]
    \centering
    \includegraphics[width=\linewidth]{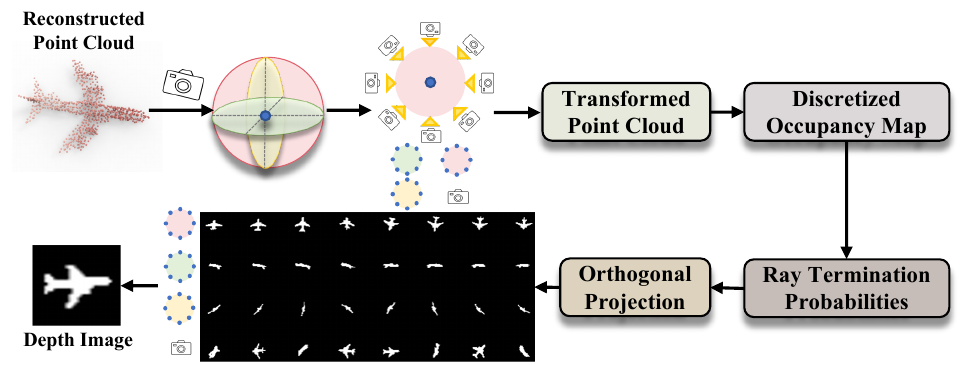}
    \vspace{-0.7cm}
    \caption{The pipeline of differentiable point cloud renderer.}
    \label{fig:dr}
\end{figure}

The rendering pipeline is illustrated in Fig.~\ref{fig:dr}, where a point-based differentiable renderer $\boldsymbol{\mathcal{R}}$~\citep{insafutdinov2018unsupervised} projects $3 \mathrm{D}$ point clouds into $2 \mathrm{D}$ images according to several camera poses.
Note that rendering views are estimated by fixing multiple camera poses rather than learning camera poses. 
The initial step entails converting the 3D coordinates of the raw point cloud into the standard coordinate frame by implementing a projective transformation that corresponds to the camera pose.
Subsequently, in order to facilitate gradient back-propagation during the training phase, the discretized point is represented through scaled Gaussian densities, thus generating an occupancy map.
The ray tracing operator, which is differentiable, transforms occupancies into probabilities of ray termination.
To obtain the projected image, the volume is projected onto the plane.
In detail, with $t$-th pose $e_t$, two types of projected images can be produced: Raw projected view images $\mathbf{I}_t=\boldsymbol{\mathcal{R}}\left(\mathcal{Q}, e_t\right)$ from ground truth $\mathcal{Q}$ and reconstructed projected view images $\hat{\mathbf{I}}_t=\boldsymbol{\mathcal{R}}\left(\hat{\mathcal{P}}, e_t\right)$ from output $\hat{\mathcal{P}}$, respectively. 
The differentiable rendering loss, denoted as $\mathcal{L}_{DR}$, is computed as mean absolute difference between the reconstructed image $\hat{\mathbf{I}}_t$, and ground truth image $\mathbf{I}_t$, for all camera poses:
\begin{equation}
    \mathcal{L}_{DR}=\frac{1}{T W H} \sum_{t=1}^T \sum_{x=1}^W \sum_{y=1}^H\left|\mathbf{I}_t(x, y)-\hat{\mathbf{I}}_t(x, y)\right|.
\end{equation}
Here, $T$ is the total number of camera poses. 
To obtain these poses, 8 cameras are evenly placed on the projection plane of each rotation axis $(\mathrm{x}, \mathrm{y}, \mathrm{z})$, where three color planes correspond to different projection planes. 
As shown in Fig.~\ref{fig:dr}, these cameras are placed at 8 diagonal positions, resulting in a total of 32 camera poses.
Each row in the planes contains the rendering images obtained from the 8 camera positions placed at the diagonal locations. 
In order to ensure the capacity of DR-Point to effectively learn the distinctive characteristics of point clouds, we integrate the differentiable rendering loss with multiple rendering images.

\subsection{Aligning Representations of Tri-Modalities}
\label{sec:Align}

Once the training triplets have been prepared, it is crucial to carefully design three branches to effectively handle the corresponding modalities. 
In particular, DR-Point implements a pre-training task to align the representations of the triplets consisting of these modalities.
The pre-training is achieved by creating a unified feature space with the help of differentiable rendering and contrastive learning. 
The learned unified feature space facilitates cross-modal applications and enhances the performance of 3D recognition in the pre-trained 3D encoder.

\subsubsection{Tri-Modal Feature Extractor}
To obtain tri-modal features, three branches are meticulously devised (Fig.~\ref{fig:overview}). 
(i) Firstly, inspired by~\citep{yu2022point}, token-level transformer autoencoder is integrated to recover point clouds at token-level, where 3D features $\mathbf{g}^P_j$ can be extracted at the same time.
In this branch, the cross-entropy loss is utilized to ensure the accurate recovery of point tokens.
(ii) Then, point-level transformer autoencoder is designed to reconstruct point clouds~\citep{liu2022masked}, which share the same encoder with the former branch.
The chamfer distance is utilized to determine the accuracy of the reconstructed 3D positions of point clouds.
Besides, to enhance the quality of the reconstructed point clouds, differentiable rendering is devised to ensure view consistency with ground truth.
Specifically, the reconstructed point clouds and their corresponding ground truth are rendered from the same camera views. This enables the calculation of a differentiable rendering loss between them. Furthermore, due to the differentiability of our devised render, it becomes possible to back-propagate the loss and update the parameters of the backbones.
Moreover, the rendered depth image can be utilized to extract depth features $\mathbf{g}^D_j$ via a ResNet~\citep{he2016deep}.
Therefore, this branch can not only promote the accuracy of point-level reconstruction but also provide depth features on the fly.
(iii) Finally, RGB image features $\mathbf{g}^R_j$ can also be obtained by applying another ResNet on the rendered RGB images.

\subsubsection{Cross-modal Contrastive Learning}
As depicted in Fig.~\ref{fig:overview}, given an object $j$, we extract RGB features $\mathbf{g}^R_j$, depth features $\mathbf{g}^D_j$, and point features $\mathbf{g}^P_j$ from the RGB, depth, and 3D point cloud branches. 
Subsequently, the contrastive loss between each pair of modalities is calculated in the following manner:
\begin{equation}
\begin{array}{r}
L_{(M_1, M_2)}=\sum_{(i, j)}-\frac{1}{2} \log \frac{\exp \left(\frac{\mathbf{g}_i^{M_1} \mathbf{g}_j^{M_2}}{\tau}\right)}{\sum_k \exp \left(\frac{\mathbf{g}_i^{M_1} \mathbf{g}_k^{M_2}}{\tau}\right)} \\
-\frac{1}{2} \log \frac{\exp \left(\frac{\mathbf{g}_i^{M_1} \mathbf{g}_j^{M_2}}{\tau}\right)}{\sum_k \exp \left(\frac{\mathbf{g}_k^{M_1} \mathbf{g}_j^{M_2}}{\tau}\right)}.
\end{array}
\end{equation}
In this equation, $M_1$ and $M_2$ correspond to two modalities, while $(i,j)$ represents a positive pair within each training batch. To introduce flexibility, we introduce a temperature parameter $\tau$, which can be learned during the optimization process.

Combing MoCo loss~\citep{he2020momentum} $\mathcal{L}_{MoCo}$ and cross-entropy loss $\mathcal{L}_{CE}$ in token-level transformer auto-encoder and $\mathcal{L}_{DR}$ and CD loss $\mathcal{L}_{CD}$ in point-level transformer auto-encoder, we minimize $L_{\text{total}}$ for all modality pairs with different coefficients,
\begin{equation}
\begin{array}{r}
    \mathcal{L}_{\text{total}}=\alpha \mathcal{L}_{(R, D)}+\beta \mathcal{L}_{(R, P)}+\theta \mathcal{L}_{(P, D)} \\ +\mathcal{L}_{MoCo} + \mathcal{L}_{CE} + \mathcal{L}_{DR} + \mathcal{L}_{CD},
\end{array}
\end{equation}
where $\alpha$, $\beta$ and $\theta$ are set to be 0.1 equally. And $\mathcal{L}_{(R, D)}, \mathcal{L}_{(R, P)}, \mathcal{L}_{(P, D)}$ represent cross-modal contrastive learning among RGB ($R$), depth ($D$), and point clouds ($P$).

\section{Pre-training Setup}
\label{sec:Setup}

\noindent\textbf{Pre-training Datasets.}
The ShapeNet dataset~\citep{chang2015shapenet} is utilized as our pre-training dataset for various point cloud understanding tasks. It encompasses over 50,000 distinct 3D models spanning 55 commonly encountered object categories.
We conducted a sampling of 1,024 points from each 3D model in ShapeNet to use as inputs. Subsequently, we divided the points into 64 groups, with each group consisting of 32 points.
Furthermore, our study incorporates a colored single-view image obtained from the ShapeNetRender dataset~\citep{afham2022crosspoint}, which serves as a valuable supplement to the ShapeNet dataset. This inclusion allows for a broader range of camera angles, enhancing the diversity of the dataset.

\noindent\textbf{Transformer Encoder.}
We intend to produce a pre-training model with a strong generalization capacity by contrastive learning of the relationship among point features, depth image features, and colored image features. 
We employ two distinct transformers: a Token-Level Transformer Auto-Encoder to acquire the point features. 
By inspiration of Point-BERT~\citep{yu2022point}, we implemented a 12-layer standard transformer encoder within the Token-Level Transformer Auto-Encoder. The hidden dimension of each encoder block was set to 384, the number of heads to 6, the FFN expansion ratio to 4, and the drop rate of stochastic depth to 0.1. 
For the Point Level Transformer Auto-Encoder, we apply the MaskTransformer~\citep{liu2022masked} to get a reconstructed point cloud and utilize the devised differentiable renderer to obtain depth images.
Then a ResNet50 is utilized to acquire the depth features.

\noindent\textbf{Token-level Transformer Auto-Encoder  Decoder.}
A single-layer Transformer decoder in token-level transformer auto-encoder is utilized for pre-training purposes. The attention block configuration is identical to that of the encoder.

\noindent\textbf{Point-level Transformer Auto-encoder  Decoder.}
The decoder of the point-level transformer auto-encoder consists of four Transformer blocks. Each of these blocks has 384 hidden dimensions and is equipped with 6 heads.

\noindent\textbf{Training Details.}
Following~\citep{yu2022point}, we conducted pre-training of DR-Point using the AdamW optimizer with a weight decay of 0.05 and a learning rate of 5 $\times 10^{-4}$, applying the cosine decay strategy. The pre-training process involved 50 epochs and a batch size of 4, with the inclusion of random scaling and translation data augmentation techniques.

\section{Downstream Task Setup}

\begin{table}[t]
\caption{\textbf{Classification on ModelNet40 dataset.} `Rep.' means we reproduce these methods.}
\centering
\tabcolsep=0.07cm
{%
\begin{tabular}{l|l|c}
\hline \toprule[1.5pt] \rowcolor{C7!50} 
                                 & Methods                  & Accuracy \\  \midrule[1pt]
\multirow{10}{*}{Supervised}     & PointNet~\cite{qi2017pointnet}                 & 89.2     \\ \hhline{~|-|-}  
                                 & \cellcolor{C7!50}PointNet++~\cite{qi2017pointnet++}             &\cellcolor{C7!50} 90.7     \\ \hhline{~|-|-} 
                                 & PointWeb~\cite{zhao2019pointweb}                 & 92.3     \\ \hhline{~|-|-} 
                                 & \cellcolor{C7!20}SpiderCNN~\cite{xu2018spidercnn}                  & \cellcolor{C7!20}92.4     \\ \hhline{~|-|-}  
                                 & PointCNN~\cite{li2018pointcnn}                 & 92.5     \\ \hhline{~|-|-} 
                                 & \cellcolor{C7!50}KPConv~\cite{thomas2019kpconv}                   & \cellcolor{C7!50}92.9     \\ \hhline{~|-|-}  
                                 & DGCNN~\cite{wang2019dynamic}                  & 92.9     \\ \hhline{~|-|-}  
                                 & \cellcolor{C7!50}RS-CNN~\cite{rao2020global}                  &\cellcolor{C7!50} 92.9     \\ \hhline{~|-|-} 
                                  & DensePoint~\cite{liu2019densepoint}                  & 93.2     \\ \hhline{~|-|-} 
                                 & \cellcolor{C7!50}PCT~\cite{guo2021pct}               & \cellcolor{C7!50}93.2     \\ \hhline{~|-|-}  
                                 & PVT~\cite{zhang2108pvt}             & 93.6     \\ \hhline{~|-|-}  
                                 & \cellcolor{C7!50}PointTransformer~\cite{zhao2021point}  & \cellcolor{C7!50}93.7     \\ \hhline{~|-|-}  
                                 & Transformer~\cite{yu2022point}     & 91.4     \\  \midrule[1pt]
\multirow{8}{*}{Self-supervised} & \cellcolor{C7!50}OcCo~\cite{wang2021unsupervised}                     & \cellcolor{C7!50}93.0     \\ \hhline{~|-|-} 
                                 & STRL~\cite{huang2021spatio}                      & 93.1     \\ \hhline{~|-|-}  
                                 & \cellcolor{C7!50} \begin{tabular}[c]{@{}c@{}} Transformer\\ +OcCo~\cite{wang2021unsupervised} \end{tabular} & \cellcolor{C7!50}92.1     \\ \hhline{~|-|-}  
                                 & Point-BERT~\cite{yu2022point}        & 93.2    \\ 
                                 \hhline{~|-|-} 
                                 & \cellcolor{C7!50}Point-MAE~\cite{pang2022masked}      & \cellcolor{C7!50}\underline{93.8}     \\
                                 \hhline{~|-|-}  
                                 & Point-MAE (Rep.)        & 93.1 \\ 
                                 \hhline{~|-|-} 
                                 & \cellcolor{C7!50}\textbf{DR-Point}      & \cellcolor{C7!50}\textbf{93.6}     \\  \bottomrule[1.5pt]
\end{tabular}%
}
\label{table:modelnet40}
\end{table}

\begin{table}[t]
\caption{\textbf{Classification on ScanObjectNN.} Accuracy ($\%$) on three settings of ScanObjectNN are listed. `Rep.' means we reproduce these methods.}
\centering
\tabcolsep=0.3cm
{%
\begin{tabular}{l|ccc}
\toprule[1.5pt]
\rowcolor{C7!50} Methods          & OBJ-BG & OBJ-ONLY & PB-T50-RS \\ \midrule[1pt]
PointNet~\cite{qi2017pointnet}           & 73.3   & 79.2     & 68.0      \\
\rowcolor{C7!50} PointNet++~\cite{qi2017pointnet++}      & 82.3   & 84.3     & 77.9      \\
DGCNN~\cite{wang2019dynamic}             & 82.8   & 86.2     & 78.1      \\
\rowcolor{C7!50}  PointCNN~\cite{li2018pointcnn}          & 86.1   & 85.5     & 78.5      \\
SpiderCNN~\cite{xu2018spidercnn}            & 77.1   & 79.5     & 73.7      \\
\rowcolor{C7!50}  BGA-DGCNN~\cite{uy2019revisiting}          & -   & -     & 79.7      \\
BGA-PN++~\cite{uy2019revisiting}             & -  & -   & 80.2      \\
\midrule[1pt]
Transformer~\cite{yu2022point}     & 79.9  & 80.6    & 77.2     \\
\rowcolor{C7!50} \begin{tabular}[c]{@{}c@{}}  Transformer\\ +OcCo~\cite{wang2021unsupervised} \end{tabular}   & 84.9  & 85.5    & 78.8     \\
Point-BERT~\cite{yu2022point}      & 87.43  & 88.12    & 83.07     \\ 
\rowcolor{C7!50}  Point-MAE~\cite{pang2022masked}          & \underline{90.02}   & \underline{88.29}     & \underline{85.18}      \\
Point-MAE (Rep.)     & 89.36  & 88.68    & 83.83     \\ 
\midrule[1pt]
\rowcolor{C7!50} \textbf{DR-Point}      & \textbf{89.51}  & \textbf{88.97}    & \textbf{84.66}     \\ \bottomrule[1.5pt]
\end{tabular}%
}
\label{table:scanobjectnn}
\end{table}

\begin{table}[ht]\small
\tabcolsep=0.1cm
\caption{\textbf{The comparison of few-shot classification performance on ModelNet40 dataset.} For a fair comparison, the average accuracy ($\%$) and standard deviation ($\%$) of 10 experiments are reported.}
\centering
{%
\begin{tabular}{l|cc|cc}
\hline \toprule[1.5pt]  
\rowcolor{C7!50}                                          & \multicolumn{2}{c|}{5-way}                   & \multicolumn{2}{c}{10-way}                   \\ \cline{2-5} \cline{2-5}
\rowcolor{C7!50}   \multirow{-2}{*}{Methods}      & \multicolumn{1}{c|}{10-shot}    & 20-shot     & \multicolumn{1}{c|}{10-shot}    & 20-shot    \\ \midrule[1pt]
DGCNN~\cite{wang2019dynamic}                                                             & \multicolumn{1}{c|}{91.8 $\pm$ 3.7} & 93.4 $\pm$ 3.2 & \multicolumn{1}{c|}{86.3 $\pm$ 6.2} & 90.9 $\pm$ 5.1 \\
\rowcolor{C7!50} \begin{tabular}[c]{@{}l@{}}DGCNN + \\ OcCo~\cite{wang2021unsupervised}\end{tabular} & \multicolumn{1}{c|}{91.9 $\pm$ 3.3} & 93.9 $\pm$ 3.1 & \multicolumn{1}{c|}{86.4 $\pm$ 5.4} & 91.3 $\pm$ 4.6 \\ \midrule[1pt]
Transformer~\cite{yu2022point}                                                     & \multicolumn{1}{c|}{87.8 $\pm$ 5.2} & 93.3 $\pm$ 4.3 & \multicolumn{1}{c|}{84.6 $\pm$ 5.5} & 89.4 $\pm$ 6.3 \\
\rowcolor{C7!50}    \begin{tabular}[c]{@{}l@{}}Transformer + \\ OcCo~\cite{wang2021unsupervised}\end{tabular} & \multicolumn{1}{c|}{94.0 $\pm$ 3.6} & 95.9 $\pm$ 2.3 & \multicolumn{1}{c|}{89.4 $\pm$ 5.1} & 92.4 $\pm$ 4.6 \\
Point-BERT~\cite{yu2022point}                                                  & \multicolumn{1}{c|}{94.6 $\pm$ 3.1} & 96.3 $\pm$ 2.7 & \multicolumn{1}{c|}{92.3 $\pm$ 4.5} & 92.7 $\pm$ 5.1 \\
\rowcolor{C7!50} MaskPoint~\cite{liu2022masked}                                                  & \multicolumn{1}{c|}{95.0 $\pm$ 3.7} & 97.2 $\pm$ 1.7 & \multicolumn{1}{c|}{91.4 $\pm$ 4.0} & 93.4 $\pm$ 3.5
\\
Point-MAE~\cite{pang2022masked}                                                  & \multicolumn{1}{c|}{96.3 $\pm$ 2.5} & 97.8 $\pm$ 1.8 & \multicolumn{1}{c|}{92.6 $\pm$ 4.1} & 95.0 $\pm$ 3.0 \\\midrule[1pt]
\rowcolor{C7!50} \textbf{DR-Point}                                                & \multicolumn{1}{c|}{\textbf{97.2 $\pm$ 2.5}} & \textbf{98.0 $\pm$ 1.8} & \textbf{93.0 $\pm$ 5.1} & \textbf{95.1 $\pm$ 3.7} \\  \bottomrule[1.5pt]
\end{tabular}%
}
\label{table:fewshot}
\end{table}

\begin{table*}[t]
\centering
\caption{\textbf{Comparison of part segmentation on ShapeNetPart dataset.} Mean IoU across all instance IoU ($\%$) is compared.}
\tabcolsep=0.6pt
\resizebox{\textwidth}{!}{%
\begin{tabular}{l|c|cccccccccccccccc}
\hline \toprule[1.5pt]
\rowcolor{C7!40} Methods          & mIoU$_I$ & Aero & Bag  & Cap  & Car  & Chair & Ear  & Guitar & Knife & Lamp & Lap  & Motor & Mug  & Pistol & Rock & Skate & table \\ \midrule[1pt]
\rowcolor{C7!50} PointNet~\cite{qi2017pointnet}          & 83.7  & 83.4 & 78.7 & 82.5 & 74.9 & 89.6  & 73.0 & 91.5   & 85.9  & 80.8 & 95.3 & 65.2  & 93.0   & 81.2   & 57.9 & 72.8  & 80.6  \\
PointNet++~\cite{qi2017pointnet++}       & 85.1  & 82.4 & 79.0   & 87.7 & 77.3 & 90.8  & 71.8 & 91.0     & 85.9  & 83.7 & 95.3 & 71.6  & 94.1 & 81.3   & 58.7 & 76.4  & \textbf{82.6}  \\
\rowcolor{C7!50} DGCNN~\cite{wang2019dynamic}            & 85.2  & 84.0   & 83.4 & 86.7 & 77.8 & 90.6  & 74.7 & 91.2   & 87.5  & 82.8 & 95.7 & 66.3  & 94.9 & 81.1   & \textbf{63.5} & 74.5  & \textbf{82.6}  \\ \midrule[1pt]
Transformer~\cite{yu2022point}       & 85.1  & 82.9 & \textbf{85.4} & 87.7 & 78.8 & 90.5  & \textbf{90.8} & 91.1   & 87.7  & 85.3 & 95.6 & 73.9  & 94.9 & 83.5   & 61.2 & 74.9  & 80.6  \\
\rowcolor{C7!50} Transformer+OcCo~\cite{wang2021unsupervised}  & 85.1  & 83.3 & 85.2 & 88.3 & \textbf{79.9} & 90.7  & 74.1 & 91.9   & 87.6  & 84.7 & 95.4 & 75.5  & 94.4 & 84.1   & 63.1 & 75.7  & 80.8  \\
Point-BERT~\cite{yu2022point}       & 85.6  & \textbf{84.3} & 84.8 & 88.0 & 79.8 & 91.0  & 81.7 & 91.6   & 87.9  & 85.2 & 95.6 & 75.6  & 94.7 & 84.3   & 63.4 & 76.3  & 81.5  \\ \midrule[1pt]
\rowcolor{C7!50}  \textbf{DR-Point}      & \textbf{86.8}  & 84.2 & 85.0  & \textbf{88.9} & 79.5  & \textbf{91.3}  & 77.0 & \textbf{92.1}   & \textbf{88.1}  &  \textbf{87.0} & \textbf{96.3} &  \textbf{76.4}   & \textbf{95.0} & \textbf{84.7} & \textbf{63.5}  & \textbf{76.9}  & 82.3  \\ \bottomrule[1.5pt]
\end{tabular}%
}
\label{table:partseg}
\end{table*}

\begin{table}[ht]
\caption{Semantic segmentation results are reported for Area 5 of the S3DIS dataset. The evaluation metrics include mAcc and mIoU across all categories. Two types of input features are employed: ``xyz'', which represents point cloud coordinates, and ``xyz+rgb'', which incorporates both coordinates and RGB color information.}
\centering
\tabcolsep=0.3cm
\begin{tabular}{l|ccc}
\toprule[1.5pt]
Methods                       & Input     & mAcc ($\%$) & mIoU ($\%$) \\ \midrule[1pt]
PointNet~\cite{qi2017pointnet}                     & xyz + rgb & 49.0      & 41.1      \\
\rowcolor{C7!50}
PointNet++~\cite{qi2017pointnet++}                    & xyz + rgb & 67.1      & 53.5      \\
PointCNN~\cite{li2018pointcnn}                      & xyz + rgb & 63.9      & 57.3      \\
\rowcolor{C7!50}
PCT~\cite{guo2021pct}                           & xyz + rgb & 67.7      & 61.3      \\ 
Transformer~\cite{yu2022point}                   & xyz       & 68.6      & 60.0      \\
\rowcolor{C7!50}
Point-BERT~\cite{yu2022point}                     & xyz       & 69.7      & 60.5      \\ 
Point-MAE~\cite{pang2022masked}                    & xyz       & 69.9      & 60.8      \\ \midrule[1pt]
\rowcolor{C7!50}
\textbf{DR-Point}              & xyz       & \textbf{70.5}         & \textbf{62.4}        \\ \bottomrule[1.5pt]
\end{tabular}
\label{tab:indoorseg}
\end{table}

\begin{table}[ht]\footnotesize
\caption{The 3D object detection results are reported on the validation set of ScanNet V2. Our pre-training model and Point-BERT adopt 3DETR as the backbone architecture. In contrast, other methods utilize VoteNet as the backbone for fine-tuning. Only geometry information is utilized as input for the downstream task. The ``Input'' column indicates the input type during the pre-training stage, where ``xyz'' represents geometry information. 
It is worth noting that the DepthContrast (xyz + rgb) model incorporates a more robust backbone (PointNet 3x) for the downstream tasks.
}
\centering
\tabcolsep=0.15cm
\begin{tabular}{l|cccc}
\toprule[1.5pt]
Methods                      & SSL & Pre-trained Input & $\textit{AP}_{25}$ & $\textit{AP}_{50}$ \\ \midrule[1pt]
\rowcolor{C7!50}
VoteNet~\cite{qi2019deep}                      &     & -                 & 58.6 & 33.5 \\
STRL~\cite{huang2021spatio}                         & \CheckmarkBold   & xyz               & 59.5 & 38.4 \\
\rowcolor{C7!50}
Implicit Autoencoder~\cite{yan2023implicit}         & \CheckmarkBold   & xyz               & 61.5 & 39.8 \\
RandomRooms~\cite{rao2021randomrooms}                  & \CheckmarkBold   & xyz               & 61.3 & 36.2 \\
\rowcolor{C7!50}
PointContrast~\cite{xie2020pointcontrast}                & \CheckmarkBold   & xyz               & 59.2 & 38.0 \\
DepthContrast~\cite{wang2021unsupervised}                & \CheckmarkBold   & xyz               & 61.3 & -    \\
\rowcolor{C7!50}
3DETR~\cite{misra2021end}                        &     & -                 & 62.1 & 37.9 \\
Point-BERT~\cite{yu2022point}                   & \CheckmarkBold   & xyz               & 61.0 & 38.3 \\
\rowcolor{C7!50}
MaskPoint~\cite{liu2022masked}                    & \CheckmarkBold   & xyz               & 63.4 & 40.6 \\
Point-MAE~\cite{pang2022masked}                    & \CheckmarkBold   & xyz               & 63.0 & 42.4 \\
\rowcolor{C7!50}
\textbf{DR-Point} & \CheckmarkBold   & xyz               & \textbf{64.0}   & \textbf{42.9}    \\
\bottomrule[1.5pt]
\end{tabular}
\label{tab:indoordet}
\end{table}



\begin{table}[ht]
\caption{Shape completion (on 16,384 points) on PCN/MVP datasets in terms of CD-$\ell_1$, CD-$\ell_2$, and F-Score@1$\%$.}
\centering
\tabcolsep=0.15cm
\begin{tabular}{l|ccc|ccc}
\toprule[1.5pt]
\rowcolor{C7!50}\multirow{2}{*}{} & \multicolumn{3}{c|}{PCN} & \multicolumn{3}{c}{MVP}  \\ \cline{2-7} 
\rowcolor{C7!50}                  & F1$\%$ & CD-$\ell_1$  & CD-$\ell_2$ & F1$\%$ & CD-$\ell_1$  & CD-$\ell_2$ \\ \midrule[1pt]
ASFMNet~\cite{xia2021asfm}           & 0.459   & 14.910 & 0.918 & 0.605   & 11.484 & 0.691 \\
\rowcolor{C7!50}GRNet~\cite{xie2020grnet}             & 0.541   & 12.790 & 0.662 & 0.609   & 11.817 & 0.679 \\
CRN~\cite{wang2021cascaded}               & 0.549   & 12.470 & 0.628 & 0.696   & 10.579 & 0.651 \\
\rowcolor{C7!50}TopNet~\cite{tchapmi2019topnet}            & 0.443   & 12.970 & 0.599 & 0.492   & 12.357 & 0.584 \\
FoldingNet~\cite{yang2018foldingnet}        & 0.418   & 12.740 & 0.570 & 0.516   & 11.881 & 0.615 \\
\rowcolor{C7!50}PCN~\cite{yuan2018pcn}              & 0.589   & 11.580 & 0.542 & 0.559   & 13.598 & 0.902 \\
ECG~\cite{pan2020ecg}               & 0.684   & 9.631  & 0.408 & 0.740   & 8.753  & 0.418 \\
\rowcolor{C7!50}PoinTr~\cite{yu2021pointr}            & 0.622   & 10.600 & 0.485 & 0.784   & 8.070  & 0.338 \\
SnowflakeNet~\cite{xiang2021snowflakenet}      & 0.743   & 8.362  & 0.311 & 0.813   & 7.597  & 0.338 \\ \midrule[1pt]
\rowcolor{C7!50} \textbf{DR-Point}          & \textbf{0.771}   & \textbf{7.478}  & \textbf{0.276} & \textbf{0.825}   & \textbf{6.473}  & \textbf{0.219} \\ \bottomrule[1.5pt]
\end{tabular}
\label{table:PCN_MVP}
\end{table}

\begin{table*}[ht]
\centering
\caption{\textbf{The comparison of DR-Point fine-tuned on ShapeNet55, ShapeNet34, and ShapeNetUnseen21 and other networks regarding CD-${\ell}_1$ {$\times 10^3$}, CD-${\ell}_2$ {$\times 10^3$} and the average F-Score@1{\%}.} Three difficult degrees including CD-\emph{S}, CD-\emph{M}, and CD-\emph{H} are leveraged to validate the completion performance, standing for the \emph{Simple}, \emph{Moderate}, and \emph{Hard} settings.}
\tabcolsep=0.6pt
\resizebox{\textwidth}{!}{%
\begin{tabular}{l|ccccc|ccccc|ccccc}
\toprule[1.5pt]
 \rowcolor{C7!40}  & \multicolumn{5}{c|}{ShapeNet55}   & \multicolumn{5}{c|}{ShapeNet34}                            & \multicolumn{5}{c}{ShapeNetUnseen21}                        \\ \cline{2-16} \cline{2-16}
 \rowcolor{C7!40}      \multirow{-2}{*}{Methods}                   & \begin{tabular}[c]{@{}c@{}}CD-\emph{S}\\ (CD-${\ell}_1$/\\ CD-${\ell}_2$)\end{tabular} & \begin{tabular}[c]{@{}c@{}}CD-\emph{M}\\ (CD-${\ell}_1$/\\ CD-${\ell}_2$)\end{tabular} & \begin{tabular}[c]{@{}c@{}}CD-\emph{H}\\ (CD-${\ell}_1$/\\ CD-${\ell}_2$)\end{tabular} & \begin{tabular}[c]{@{}c@{}}CD-\emph{Avg.}\\ (CD-${\ell}_1$/\\ CD-${\ell}_2$)\end{tabular} & \begin{tabular}[c]{@{}c@{}}F-Socre\\ -Avg\end{tabular} & \begin{tabular}[c]{@{}c@{}}CD-\emph{S}\\ (CD-${\ell}_1$/\\ CD-${\ell}_2$)\end{tabular} & \begin{tabular}[c]{@{}c@{}}CD-\emph{M}\\ (CD-${\ell}_1$/\\ CD-${\ell}_2$)\end{tabular} & \begin{tabular}[c]{@{}c@{}}CD-\emph{H}\\ (CD-${\ell}_1$/\\ CD-${\ell}_2$)\end{tabular} & \begin{tabular}[c]{@{}c@{}}CD-\emph{Avg.}\\ (CD-${\ell}_1$/\\ CD-${\ell}_2$)\end{tabular} & \begin{tabular}[c]{@{}c@{}}F-Socre\\ -Avg\end{tabular} & \begin{tabular}[c]{@{}c@{}}CD-\emph{S}\\ (CD-${\ell}_1$/\\ CD-${\ell}_2$)\end{tabular} & \begin{tabular}[c]{@{}c@{}}CD-\emph{M}\\ (CD-${\ell}_1$/\\ CD-${\ell}_2$)\end{tabular} & \begin{tabular}[c]{@{}c@{}}CD-\emph{H}\\ (CD-${\ell}_1$/\\ CD-${\ell}_2$)\end{tabular} & \begin{tabular}[c]{@{}c@{}}CD-\emph{Avg.}\\ (CD-${\ell}_1$/\\ CD-${\ell}_2$)\end{tabular} & \begin{tabular}[c]{@{}c@{}}F-Socre\\ -Avg\end{tabular} \\ \midrule[1pt]
\rowcolor[HTML]{FFFFFF} ASFMNet~\cite{xia2021asfm}                  & \begin{tabular}[c]{@{}c@{}}19.138\\ 1.308\end{tabular}          & \begin{tabular}[c]{@{}c@{}}20.172\\ 1.517\end{tabular}          & \begin{tabular}[c]{@{}c@{}}23.513\\ 2.282\end{tabular}          & \begin{tabular}[c]{@{}c@{}}20.941\\ 1.702\end{tabular}            & 0.247                                                  & \begin{tabular}[c]{@{}c@{}}18.350\\ 1.189\end{tabular}          & \begin{tabular}[c]{@{}c@{}}19.123\\ 1.343\end{tabular}          & \begin{tabular}[c]{@{}c@{}}21.913\\ 1.909\end{tabular}          & \begin{tabular}[c]{@{}c@{}}19.795\\ 1.480\end{tabular}            & 0.268                                                  & \begin{tabular}[c]{@{}c@{}}21.591\\ 1.995\end{tabular}          & \begin{tabular}[c]{@{}c@{}}23.006\\ 2.342\end{tabular}          & \begin{tabular}[c]{@{}c@{}}27.682\\ 3.660\end{tabular}          & \begin{tabular}[c]{@{}c@{}}24.075\\ 2.666\end{tabular}            & 0.216                                                  \\
\rowcolor{C7!50} TopNet~\cite{tchapmi2019topnet}                   & \begin{tabular}[c]{@{}c@{}}27.233\\ 2.483\end{tabular}          & \begin{tabular}[c]{@{}c@{}}28.749\\ 2.848\end{tabular}          & \begin{tabular}[c]{@{}c@{}}33.986\\ 4.642\end{tabular}          & \begin{tabular}[c]{@{}c@{}}29.989\\ 3.324\end{tabular}            & 0.110                                                  & \begin{tabular}[c]{@{}c@{}}22.382\\ 1.606\end{tabular}          & \begin{tabular}[c]{@{}c@{}}23.271\\ 1.793\end{tabular}          & \begin{tabular}[c]{@{}c@{}}26.020\\ 2.432\end{tabular}          & \begin{tabular}[c]{@{}c@{}}23.891\\ 1.944\end{tabular}            & 0.154                                                  & \begin{tabular}[c]{@{}c@{}}26.775\\ 2.499\end{tabular}          & \begin{tabular}[c]{@{}c@{}}28.312\\ 2.928\end{tabular}          & \begin{tabular}[c]{@{}c@{}}33.121\\ 4.407\end{tabular}          & \begin{tabular}[c]{@{}c@{}}29.403\\ 3.278\end{tabular}            & 0.103                                                  \\
\rowcolor[HTML]{FFFFFF} GRNet~\cite{xie2020grnet}                    & \begin{tabular}[c]{@{}c@{}}19.159\\ 1.137\end{tabular}          & \begin{tabular}[c]{@{}c@{}}20.645\\ 1.489\end{tabular}          & \begin{tabular}[c]{@{}c@{}}24.034\\ 2.394\end{tabular}          & \begin{tabular}[c]{@{}c@{}}21.279\\ 1.673\end{tabular}            & 0.239                                                  & \begin{tabular}[c]{@{}c@{}}18.809\\ 1.102\end{tabular}          & \begin{tabular}[c]{@{}c@{}}20.034\\ 1.366\end{tabular}          & \begin{tabular}[c]{@{}c@{}}22.989\\ 2.089\end{tabular}          & \begin{tabular}[c]{@{}c@{}}20.611\\ 1.519\end{tabular}            & 0.247                                                  & \begin{tabular}[c]{@{}c@{}}21.245\\ 1.552\end{tabular}          & \begin{tabular}[c]{@{}c@{}}23.753\\ 2.281\end{tabular}          & \begin{tabular}[c]{@{}c@{}}49.427\\ 4.169\end{tabular}          & \begin{tabular}[c]{@{}c@{}}24.808\\ 2.667\end{tabular}            & 0.208                                                  \\
\rowcolor{C7!50}  FoldingNet~\cite{yang2018foldingnet}               
& \begin{tabular}[c]{@{}c@{}}25.203\\ 2.095\end{tabular}          & \begin{tabular}[c]{@{}c@{}}26.596\\ 2.410\end{tabular}          & \begin{tabular}[c]{@{}c@{}}30.424\\ 3.333\end{tabular}          & \begin{tabular}[c]{@{}c@{}}27.408\\ 2.613\end{tabular}            & 0.091                                                  & \begin{tabular}[c]{@{}c@{}}23.556\\ 1.859\end{tabular}          & \begin{tabular}[c]{@{}c@{}}24.466\\ 2.059\end{tabular}          & \begin{tabular}[c]{@{}c@{}}27.584\\ 2.759\end{tabular}          & \begin{tabular}[c]{@{}c@{}}25.202\\ 2.226\end{tabular}            & 0.137                                                  & \begin{tabular}[c]{@{}c@{}}28.356\\ 2.887\end{tabular}          & \begin{tabular}[c]{@{}c@{}}29.833\\ 3.290\end{tabular}          & \begin{tabular}[c]{@{}c@{}}35.356\\ 4.968\end{tabular}          & \begin{tabular}[c]{@{}c@{}}31.182\\ 3.715\end{tabular}            & 0.088                                                  \\
\rowcolor[HTML]{FFFFFF} CRN~\cite{wang2021cascaded}                     
& \begin{tabular}[c]{@{}c@{}}21.207\\ 1.502\end{tabular}          & \begin{tabular}[c]{@{}c@{}}22.364\\ 1.801\end{tabular}          & \begin{tabular}[c]{@{}c@{}}25.849\\ 2.726\end{tabular}          & \begin{tabular}[c]{@{}c@{}}23.140\\ 2.010\end{tabular}            & 0.205                                                  & \begin{tabular}[c]{@{}c@{}}20.304\\ 1.362\end{tabular}          & \begin{tabular}[c]{@{}c@{}}21.216\\ 1.594\end{tabular}          & \begin{tabular}[c]{@{}c@{}}24.159\\ 2.318\end{tabular}          & \begin{tabular}[c]{@{}c@{}}21.893\\ 1.758\end{tabular}            & 0.221                                                  & \begin{tabular}[c]{@{}c@{}}24.247\\ 2.237\end{tabular}          & \begin{tabular}[c]{@{}c@{}}26.076\\ 2.840\end{tabular}          & \begin{tabular}[c]{@{}c@{}}31.771\\ 4.833\end{tabular}          & \begin{tabular}[c]{@{}c@{}}27.365\\ 3.303\end{tabular}            & 0.177                                                  \\
\rowcolor{C7!50} PCN~\cite{yuan2018pcn} & \begin{tabular}[c]{@{}c@{}}22.990\\ 1.811\end{tabular}          & \begin{tabular}[c]{@{}c@{}}23.976\\ 2.062\end{tabular}          & \begin{tabular}[c]{@{}c@{}}27.360\\ 2.937\end{tabular}          & \begin{tabular}[c]{@{}c@{}}24.775\\ 2.270\end{tabular}            & 0.167                                                  & \begin{tabular}[c]{@{}c@{}}21.433\\ 1.551\end{tabular}          & \begin{tabular}[c]{@{}c@{}}22.304\\ 1.753\end{tabular}          & \begin{tabular}[c]{@{}c@{}}25.086\\ 2.426\end{tabular}          & \begin{tabular}[c]{@{}c@{}}22.941\\ 1.910\end{tabular}            & 0.192                                                  & \begin{tabular}[c]{@{}c@{}}27.593\\ 2.983\end{tabular}          & \begin{tabular}[c]{@{}c@{}}28.989\\ 3.442\end{tabular}          & \begin{tabular}[c]{@{}c@{}}34.598\\ 5.558\end{tabular}          & \begin{tabular}[c]{@{}c@{}}30.393\\ 3.994\end{tabular}            & 0.128                                                  \\
\rowcolor[HTML]{FFFFFF} ECG~\cite{pan2020ecg}                     
& \begin{tabular}[c]{@{}c@{}}16.710\\ 1.167\end{tabular}          & \begin{tabular}[c]{@{}c@{}}18.727\\ 1.545\end{tabular}          & \begin{tabular}[c]{@{}c@{}}23.480\\ 2.555\end{tabular}          & \begin{tabular}[c]{@{}c@{}}19.639\\ 1.756\end{tabular}            & 0.321                                                  & \begin{tabular}[c]{@{}c@{}}13.122\\ 0.735\end{tabular}          & \begin{tabular}[c]{@{}c@{}}14.628\\ 0.996\end{tabular}          & \begin{tabular}[c]{@{}c@{}}18.461\\ 1.696\end{tabular}          & \begin{tabular}[c]{@{}c@{}}15.404\\ 1.142\end{tabular}            & \textbf{0.496}                                                  & \begin{tabular}[c]{@{}c@{}}15.282\\ 1.255\end{tabular}          & \begin{tabular}[c]{@{}c@{}}17.595\\ 1.759\end{tabular}          & \begin{tabular}[c]{@{}c@{}}23.535\\ 3.267\end{tabular}          & \begin{tabular}[c]{@{}c@{}}18.804\\ 2.094\end{tabular}            & \textbf{0.460}                                                  \\
\rowcolor{C7!50} PoinTr~\cite{yu2021pointr}                 & \begin{tabular}[c]{@{}c@{}}12.491\\ 0.698\end{tabular}          & \begin{tabular}[c]{@{}c@{}}14.182\\ 1.049\end{tabular}          & \begin{tabular}[c]{@{}c@{}}18.811\\ 2.022\end{tabular}          & \begin{tabular}[c]{@{}c@{}}15.161\\ 1.256\end{tabular}            & \textbf{0.446}                                                  & \begin{tabular}[c]{@{}c@{}}12.006\\ 0.632\end{tabular}          & \begin{tabular}[c]{@{}c@{}}13.393\\ 0.910\end{tabular}          & \begin{tabular}[c]{@{}c@{}}17.365\\ 1.697\end{tabular}          & \begin{tabular}[c]{@{}c@{}}14.255\\ 1.080\end{tabular}            & 0.459                                                  & \begin{tabular}[c]{@{}c@{}}13.290\\ 0.838\end{tabular}          & \begin{tabular}[c]{@{}c@{}}15.522\\ 1.376\end{tabular}          & \begin{tabular}[c]{@{}c@{}}21.881\\ 3.070\end{tabular}          & \begin{tabular}[c]{@{}c@{}}16.898\\ 1.761\end{tabular}            & 0.421                                                  \\
\rowcolor[HTML]{FFFFFF} SnowflakeNet~\cite{xiang2021snowflakenet}             & \begin{tabular}[c]{@{}c@{}}13.568\\ 0.680\end{tabular}          & \begin{tabular}[c]{@{}c@{}}15.380\\ 0.979\end{tabular}          & \begin{tabular}[c]{@{}c@{}}19.412\\ \textbf{1.754}\end{tabular}          & \begin{tabular}[c]{@{}c@{}}16.120\\ 1.138\end{tabular}            & 0.362                                                  & \begin{tabular}[c]{@{}c@{}}13.612\\ 0.693\end{tabular}          & \begin{tabular}[c]{@{}c@{}}15.272\\ 0.968\end{tabular}          & \begin{tabular}[c]{@{}c@{}}19.385\\ 1.727\end{tabular}          & \begin{tabular}[c]{@{}c@{}}16.090\\ 1.129\end{tabular}            & 0.370                                                  & \begin{tabular}[c]{@{}c@{}}15.162\\ 0.974\end{tabular}          & \begin{tabular}[c]{@{}c@{}}17.720\\ 1.491\end{tabular}          & \begin{tabular}[c]{@{}c@{}}23.986\\ 3.022\end{tabular}          & \begin{tabular}[c]{@{}c@{}}18.956\\ 1.829\end{tabular}            & 0.331                                                  \\ \midrule[1pt]
\rowcolor{C7!50}  \textbf{DR-Point}              
& \begin{tabular}[c]{@{}c@{}} \textbf{10.089}\\ \textbf{0.572}\end{tabular}          & \begin{tabular}[c]{@{}c@{}}\textbf{11.904}\\ \textbf{0.931}\end{tabular}          & \begin{tabular}[c]{@{}c@{}}\textbf{16.135}\\ 1.875 \end{tabular}          & \begin{tabular}[c]{@{}c@{}}\textbf{12.709}\\ \textbf{1.126} \end{tabular}            &  0.415                                               & \begin{tabular}[c]{@{}c@{}}\textbf{9.819}\\ \textbf{0.535}\end{tabular}           & \begin{tabular}[c]{@{}c@{}}\textbf{11.364}\\ \textbf{0.818}\end{tabular}          & \begin{tabular}[c]{@{}c@{}}\textbf{15.056} \\ \textbf{1.595}\end{tabular}          & \begin{tabular}[c]{@{}c@{}}\textbf{12.080}\\\textbf{0.983} \end{tabular}            &    0.431                              & \begin{tabular}[c]{@{}c@{}}\textbf{10.496} \\ \textbf{0.673} \end{tabular}          & \begin{tabular}[c]{@{}c@{}}\textbf{12.749} \\ \textbf{1.158}\end{tabular}          & \begin{tabular}[c]{@{}c@{}}\textbf{17.947} \\ \textbf{2.427} \end{tabular}          & \begin{tabular}[c]{@{}c@{}}\textbf{13.731}\\ \textbf{1.419}\end{tabular}            &   0.400                                              \\ \bottomrule[1.5pt]
\end{tabular}%
}
\label{table:completion-ShapeNet}
\end{table*}

\begin{figure}[t]
    \centering
    \includegraphics[width=\linewidth]{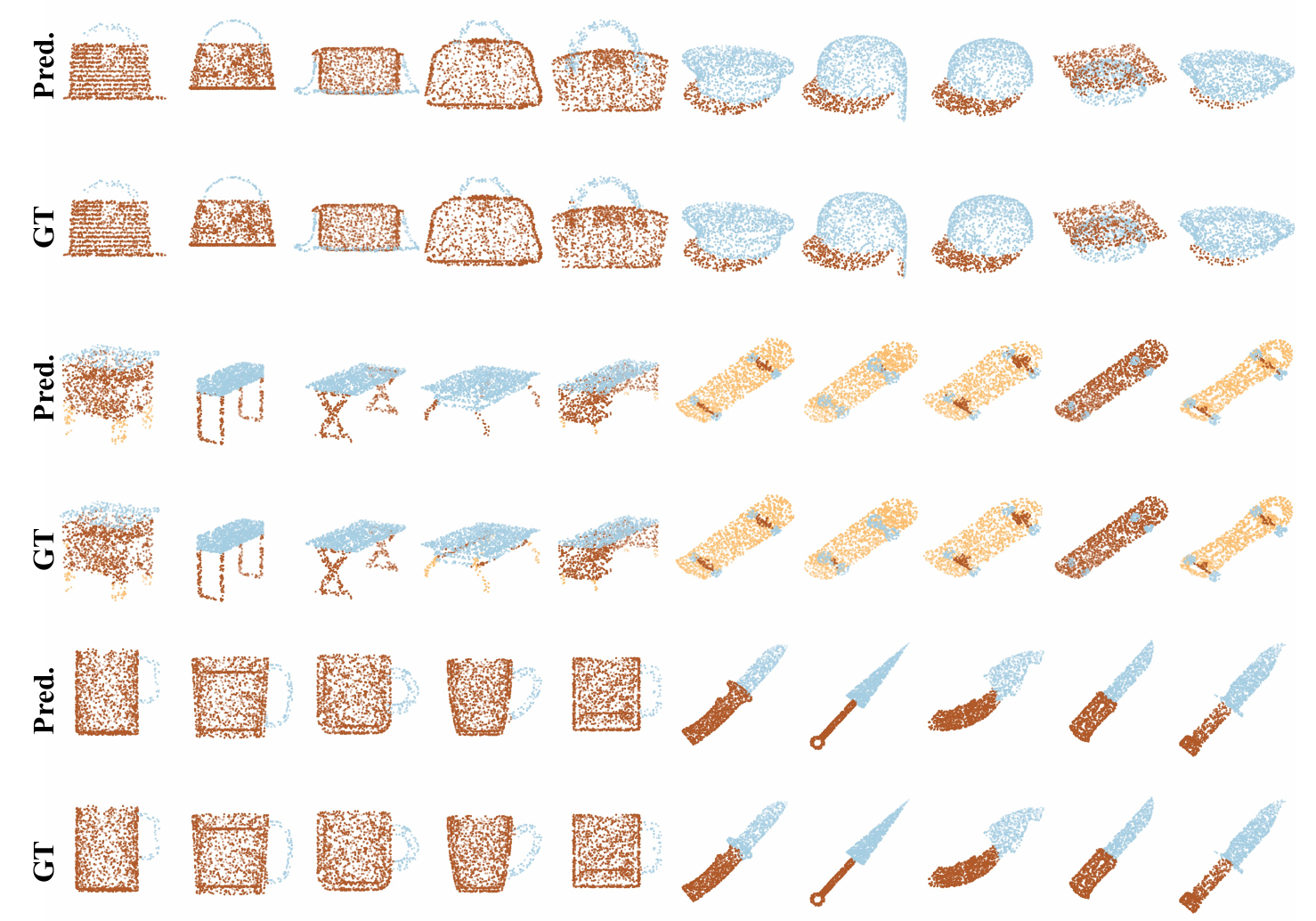}
    \vspace{-0.7cm}
    \caption{Visualization comparison of segmentation on ShapeNetPart.}
    \label{fig:seg}
\end{figure}


\begin{figure*}[t]
    \centering
    \includegraphics[width=\linewidth]{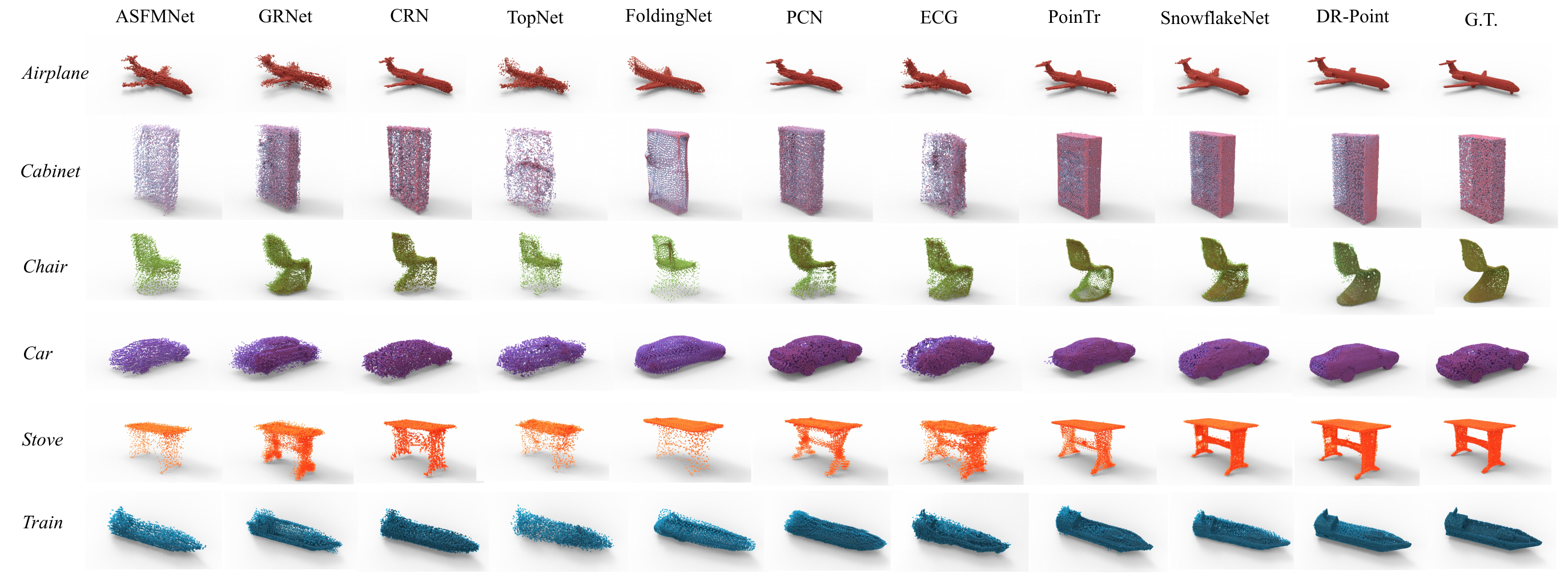}
    \vspace{-0.8cm}
    \caption{Visualization comparisons on PCN dataset, which is the commonly used point cloud completion dataset.}
    \label{fig:PCN_vis_results}
\end{figure*}

\begin{figure*}[ht]
    \centering
    \includegraphics[width=\linewidth]{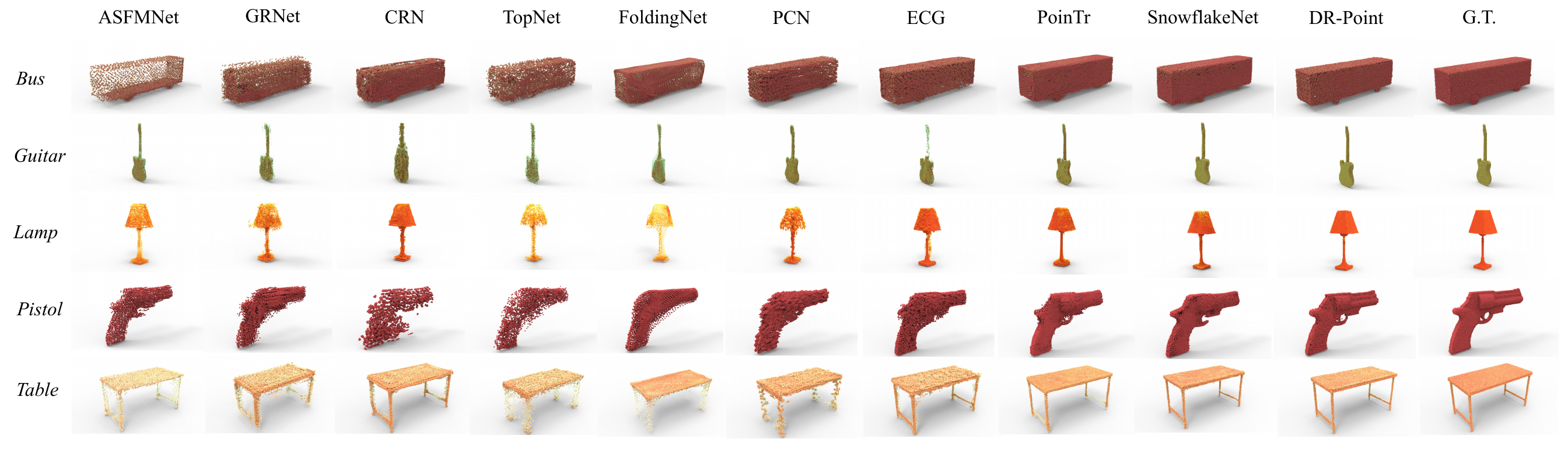}
    \vspace{-0.7cm}
    \caption{Visualization comparisons on MVP dataset, containing various incomplete patterns.}
    \label{fig:MVP_vis_results}
\end{figure*}

\begin{figure*}[ht]
    \centering
    \includegraphics[width=\linewidth]{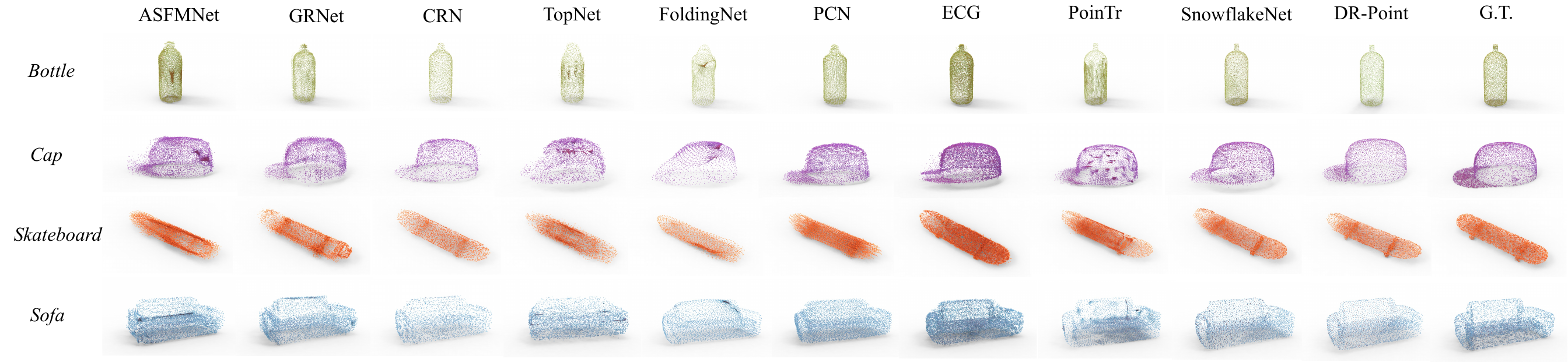}
    \vspace{-0.7cm}
    \caption{Visualization comparisons on ShapeNet55 dataset, which utilizes all categories of ShapeNet.}
    \label{fig:ShapeNet55_vis_results}
\end{figure*}

\begin{figure*}[ht]
    \centering
    \includegraphics[width=\linewidth]{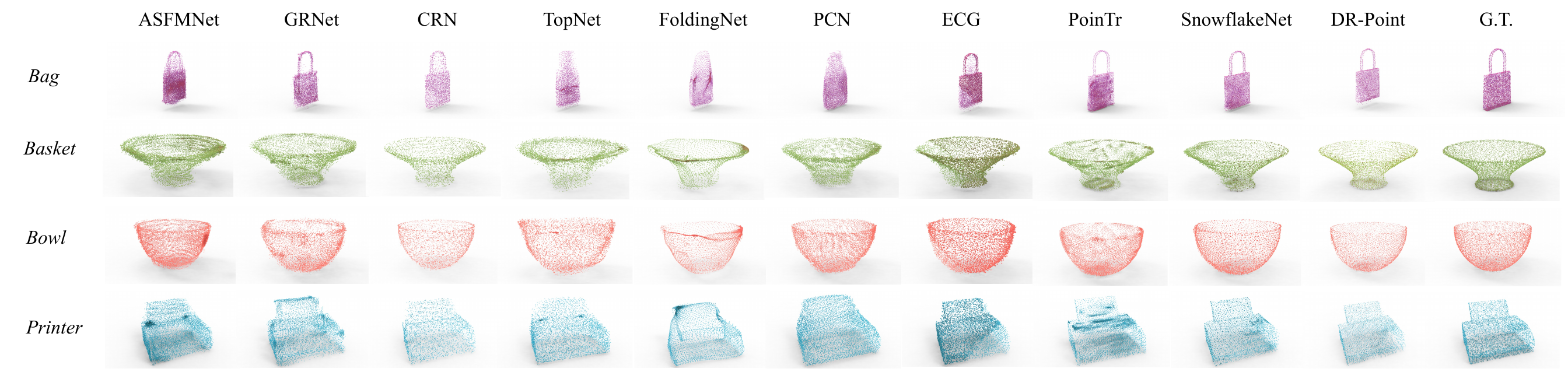}
    \vspace{-0.7cm}
    \caption{Visualization comparisons on ShapeNetUnseen21 dataset, which is utilized to validate the generalize capabilities.}
    \label{fig:Unseen21_vis_results}
\end{figure*}

\begin{table*}[t]
\tabcolsep=0.32cm
\centering
\caption{Ablation studies on the weight factors of losses.}
\resizebox{\textwidth}{!}{
\begin{tabular}{c|ccccccc|cc}
\toprule[1.5pt]
Model    & $\mathcal{L}_{(R, D)}$ & $\mathcal{L}_{(R, P)}$  & $\mathcal{L}_{(P, D)}$ & MoCo & CE & DR & CD & ModelNet40 ($\%$) & ScanObjectNN (OBJ-BG) ($\%$) \\ \midrule[1pt]
\rowcolor{C7!50} A        & 0.01 & 0.01 & 0.01 & 1    & 1  & 1  & 1  & 92.65     & 88.34     \\
B        & 0.5  & 0.5  & 0.5  & 1    & 1  & 1  & 1  &  92.82    &  88.52    \\
\rowcolor{C7!50} C        & 1    & 1    & 1    & 1    & 1  & 1  & 1  & 92.57     &  88.28 \\  
D        & 0.1    & 0.1    & 0.1    & 0    & 1  & 1  & 1  & 93.01     & 88.98 \\
\rowcolor{C7!50} E        & 0.1    & 0.1    & 0.1    & 2    & 1  & 1  & 1  & 93.55     & 89.43 \\
F        & 0.1    & 0.1    & 0.1    & 1    & 1  & 0  & 1  & 92.88     & 88.65 \\ \midrule[0.8pt]
\rowcolor{C7!50} DR-Point & 0.1  & 0.1  & 0.1  & 1    & 1  & 1  & 1  &  \textbf{93.60}    & \textbf{89.51}  \\  \bottomrule[1.5pt]
\end{tabular}
}
\label{r1}
\end{table*}

\begin{table}[t]\small
\caption{\textbf{Ablation studies} are conducted on ModelNet40 and ScanObjectNN (OBJ-BG) to evaluate the alignments between different modalities.}
\centering
\setlength\tabcolsep{1.5pt}
\begin{tabular}{l|ccccc}
\toprule[1.5pt]   \rowcolor{C7!50}       & RD-CL & RP-CL & DP-CL & ModelNet40 & ScanObjectNN \\ \midrule[1pt]
Model A &       & \CheckmarkBold     &       &  92.4                 &       88.3               \\ 
\rowcolor{C7!50} Model B &       &       & \CheckmarkBold     &   92.2                 &    88.5                  \\ 
Model C &       & \CheckmarkBold     & \CheckmarkBold     &    93.3                &     89.1                 \\ 
\rowcolor{C7!50} Model D & \CheckmarkBold     &       & \CheckmarkBold     &         93.1           &     88.7               \\ 
Model E & \CheckmarkBold     & \CheckmarkBold     &       &    93.0                &     88.9                 \\ \midrule[1pt]
\rowcolor{C7!50} \textbf{DR-Point} & \CheckmarkBold     & \CheckmarkBold     & \CheckmarkBold     &     \textbf{93.6}               &    \textbf{89.5}                   \\ \bottomrule[1.5pt]
\end{tabular}
\label{ablation}
\end{table}

\begin{table}[t]\small
\caption{\textbf{Ablation studies} are conducted on ModelNet40 to evaluate the influence of RGB and depth images.}
\centering
\begin{tabular}{l|cccccc}
\toprule[1.5pt]
\rowcolor{C7!50} \begin{tabular}[c]{@{}l@{}}Number of \\ RGB Images\end{tabular}   & 1 & 2 & 3 & 4 & 5 & 6 \\ \midrule[1pt]
  \begin{tabular}[c]{@{}l@{}}Number of \\ Depth Images\end{tabular} & 1 & 2 & 3 & 4 & 5 & 6 \\ \midrule[1pt]
\rowcolor{C7!50}ModelNet40                                                        & \textbf{93.6}  & 93.5  & 93.2  &  93.0 & 93.1  &  92.8 \\ \bottomrule[1.5pt]
\end{tabular}
\label{ablation-images}
\end{table}

\begin{table}[t]
\caption{Ablation study on the rendered images for enhancing the accuracy of the reconstruction and downstream tasks.}
\centering
\begin{tabular}{l|llll}
\toprule[1.5pt]
Number of Depth Images & 8    & 16   & 24   & 32   \\ \midrule[1pt]
Acc. on ModelNet40     & 92.7 & 93.3 & 93.4 & \textbf{93.6} \\ \bottomrule[1.5pt]
\end{tabular}
\label{ablation3}
\end{table}

\begin{figure}[t]
    \centering
    \includegraphics[width=0.95\linewidth]{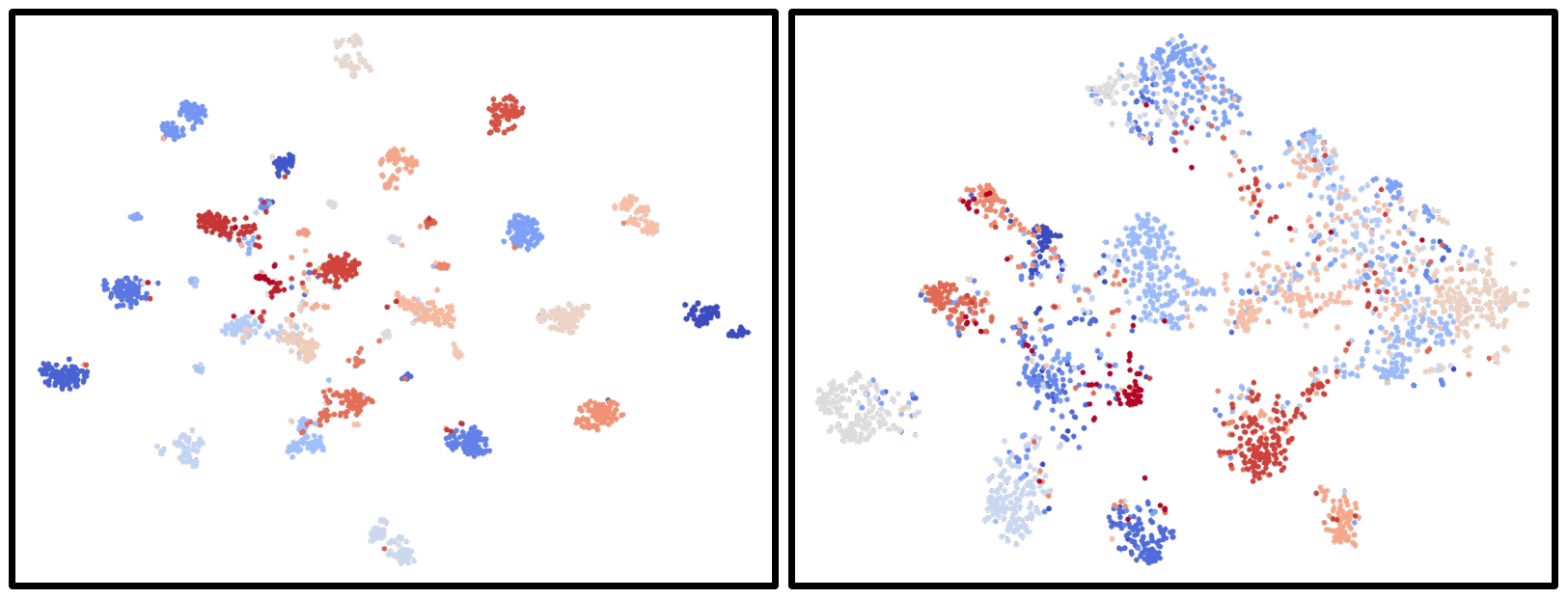}
    \vspace{-0.3cm}
    \caption{Visualization of feature distributions of our DR-Point after fine-tuning on ModelNet40 (\textbf{left}) and ScanObjectNN (\textbf{right}).}
    \label{fig:tsne}
\end{figure}


\noindent\textbf{Shape Classification.} 
We conducted experiments on two benchmarks, namely ModelNet40~\citep{wu20153d} and ScanObjectNN~\citep{uy2019revisiting}, to evaluate the effectiveness of our object classification method. 
Synthetic object classification was performed on ModelNet40, while real-world object classification was conducted on ScanObjectNN. To ensure consistency, we adopted the same settings as~\citep{qi2017pointnet,qi2017pointnet++} for fine-tuning. All models were trained for 200 epochs with a batch size of 32.

\noindent\textbf{Few-shot Classification.} 
In accordance with previous studies~\citep{wang2021unsupervised, zhang2021self, wang2021unsupervised}, we apply the ``$K$-way $N$-shot'' approach to conduct few-shot classification on the ModelNet40 dataset~\citep{yu2022point}.
Specifically, we randomly select $K$ out of the 40 available categories and $N$+20 3D objects per category, where $N$ objects are used for training and 20 objects for testing.
The DR-Point is evaluated on four few-shot scenarios: 5-way 10-shot, 5-way 20-shot, 10-way 10-shot, and 10-way 20-shot, respectively. Further, 10 independent runs under each setting are performed, and average accuracy together with standard deviations are reported to minimize the influence of the variance of random sampling. The fine-tuning settings are still just same as 3D shape classification, but the epochs decrease to 150 epochs.

\noindent\textbf{Part Segmentation.} 
For the task of fine-grained 3D recognition, specifically part segmentation, we utilize ShapeNetPart~\citep{yi2016scalable}, a comprehensive dataset consisting of 16,881 objects. Each object is represented by 2,048 points and belongs to one of 16 categories, with a total of 50 distinct parts.
Similar to PointNet~\citep{qi2017pointnet}, we conducted a sampling of 2,048 points from each model. The models were trained over 250 epochs, utilizing a batch size of 16.

\noindent\textbf{Point Cloud Completion.} 
In order to tackle the point cloud completion task, we employ a conventional Transformer encoder alongside a robust Transformer-based decoder, as proposed in the SnowflakeNet architecture by~\citep{xiang2021snowflakenet}.
Our model is fine-tuned on the point cloud completion benchmarks, undergoing 200 epochs of training.

\noindent\textbf{Indoor Segmentation.}
Consistent with established conventions, we designated area 5 of S3DIS specifically for testing purposes, while utilizing the remaining areas for training our models.

\noindent\textbf{Indoor Detection.}
We adopt the evaluation procedure established by VoteNet~\citep{qi2019deep}, which calculates the mean average precision for two threshold values: 0.25 (mAP@0.25) and 0.5 (mAP@0.5). These metrics allow us to effectively evaluate the performance of our DR-Point.

\section{Dataset Briefs}
\label{sec:Dataset}

\textbf{ModelNet40~\citep{wu20153d}}: 
The ModelNet40 dataset is a collection of synthetic object point clouds commonly used as a benchmark for point cloud analysis tasks. It is popular due to its diverse range of object categories, clean and well-defined shapes, and carefully constructed dataset. The original ModelNet40 dataset consists of 12,311 computer-aided design (CAD) generated meshes representing objects from 40 categories such as airplanes, cars, plants, lamps, and more. For training and testing purposes, the dataset is split into two sets: a training set (which contains 9,843 CAD-generated meshes) and a testing set (which includes the remaining 2,468 meshes). 

\noindent\textbf{ScanObjectNN~\citep{uy2019revisiting}}: 
The ScanObjectNN dataset comprises around 15,000 meticulously classified objects, divided into 15 distinct categories. It contains a total of 2,902 unique instances of objects. Each object in the dataset is represented by a comprehensive set of attributes, including a list of points with both global and local coordinates, corresponding normals, color information, and semantic labels.

\noindent\textbf{ShapeNetPart~\citep{yi2016scalable}}: 
The ShapeNetPart dataset is an extension of the original ShapeNet~\citep{yu2021pointr} dataset, providing detailed part-level annotations for different classes of objects, specifically designed for part-level semantic segmentation tasks in 3D shape analysis. The dataset contains models of different classes of 3D objects, including everyday objects, furniture, vehicles, and so on. 

\noindent\textbf{PCN dataset~\citep{yuan2018pcn}}: 
The PCN dataset serves as a widely used benchmark dataset for point cloud completion tasks. However, it is limited to only eight categories derived from the ShapeNet dataset. In the PCN dataset, incomplete shapes are created by projecting complete shapes from eight distinct viewpoints. Each complete point cloud within the dataset comprises a total of 16,384 points.

\noindent\textbf{MVP~\citep{pan2021variational}}: MVP dataset\citep{pan2021variational} expands the existing 8 categories in the PCN dataset by introducing an additional 8 categories, including bed, bench, bookshelf, bus, guitar, motorbike, pistol, and skateboard, resulting in a comprehensive set of high-quality partial and complete point clouds.

\noindent\textbf{ShapeNet55/34~\citep{yu2021pointr}}: Traditionally, point cloud completion datasets (e.g., PCN~\citep{yuan2018pcn}) focused on limited categories, disregarding the diversity of real-world uncompleted point clouds. To overcome this, the ShapeNet55 benchmark leverages objects from 55 categories, enabling a thorough evaluation of model capabilities. ShapeNet34/ShapeNet Unseen21 split the original dataset into two parts: 34 seen categories used for training and 21 unseen categories. This division evaluates models’ generalization to handle unseen categories based on knowledge from seen categories. These benchmarks provide valuable insights into point cloud completion models' performance across object categories, fostering the development of robust models for real-world challenges.

\noindent\textbf{Indoor Segmentation.}
The S3DIS dataset, commonly referred to as the Stanford Large-Scale 3D Indoor Spaces dataset~\citep{armeni20163d}, provides instance-level semantic segmentation for six large indoor areas.
These areas consist of a total of 271 rooms and encompass 13 distinct semantic categories. 
Consistent with established conventions, we designated area 5 specifically for testing purposes, while utilizing the remaining areas for training our models.

\noindent\textbf{Indoor Detection.}
The benchmark widely recognized for 3D object detection is ScanNet V2~\citep{dai2017scannet}, which comprises 1,513 indoor scenes and encompasses 18 distinct object classes. To ensure consistency, we adopt the evaluation procedure established by VoteNet~\citep{qi2019deep}, which calculates the mean average precision for two threshold values: 0.25 ($\textit{AP}_{25}$) and 0.5 ($\textit{AP}_{50}$). These metrics allow us to effectively evaluate the performance of our DR-Point.

\section{Experiments}
\label{sec:Experiments}

Within this section, we conduct an assessment of DR-Point's performance across a range of downstream tasks. These tasks encompass shape classification, few-shot classification, part segmentation, point cloud completion, semantic segmentation, and detection. 

\noindent\subsection{Object Classification on Clean Shapes} 
As shown in Table~\ref{table:modelnet40}, DR-Point achieves a remarkable overall accuracy (OA) improvement of 2.7$\%$ with 1k points compared to Transformer trained from scratch. 
Moreover, it produces a gain of 1.9$\%$ over OcCo~\citep{wang2021unsupervised} pre-training and 0.8$\%$ over Point-BERT~\citep{yu2022point} pre-training. 
This considerable improvement over the baselines demonstrates the effectiveness of our pre-training methodology. 
Significantly, our standard vision transformer architecture achieves comparable performance to the intricately designed attention operators from PointTransformer~\citep{zhao2021point}, when evaluated with 1k points (93.6$\%$ vs 93.7$\%$).

\noindent\subsection{Object Classification on Real-World Dataset} 
Moreover, we performed experiments on three distinct variants of ScanObjectNN~\citep{uy2019revisiting}, specifically referred to as \textit{OBJ-BG}, \textit{OBJ ONLY}, and \textit{PB-T50-RS}. The outcomes of these experiments are illustrated in Table~\ref{table:scanobjectnn}. 
Our DR-Point significantly improves the baseline performance by 12.0$\%$, 10.3$\%$, and 9.6$\%$ for the three variants correspondingly. 
Particularly on the most challenging variant \textit{PB-T50-RS}, our proposed model achieved an accuracy of 84.6$\%$, which outperformed Point-BERT~\citep{yu2022point} by 1.9$\%$. 
Remarkably, despite being pre-trained on images of clean objects, our DR-Point exhibits remarkable generalization ability on real-world data, showcasing its impressive capability to generalize effectively.

\noindent\subsection{Few-shot Object Classification} 
In order to assess the few-shot classification performance of DR-Point with limited fine-tuning data, we carried out experiments on Few-shot ModelNet40.
DR-Point's superior performance is supported by Table~\ref{table:fewshot}, with superiority of +0.9$\%$, +0.2$\%$, +0.4$\%$, and +0.1$\%$, respectively, over Point-MAE in all four settings. 
Moreover, smaller deviations were observed with our approach compared to other transformer-based methods, which suggests our DR-Point produces more universally adaptable 3D representations in low-data regimes.

\noindent\subsection{3D Object Part Segmentation}
Table~\ref{table:partseg} presents the results for 3D Object Part Segmentation. The DR-Point method demonstrates superior performance compared to the training from scratch approach (PointViT) and the OcCo-pretraining baseline. 
Furthermore, it achieves a 1.4$\%$ improvement in comparison to Point-BERT. 
This notable performance enhancement is attributed to our tri-modal pre-training objective, which involves the dense classification of points throughout the 3D space.
Consequently, we achieve outstanding results when scaling up to dense prediction tasks.

\noindent\subsection{Indoor 3D Semantic Segmentation} 
Moreover, our study aims to assess the effectiveness of the DR-Point in the context of 3D semantic segmentation for large-scale scenes. This particular task presents notable difficulties, as it necessitates comprehending both the overall semantic context and the intricate geometric details at a local level. 
The outcomes of our experiments are outlined in Table~\ref{tab:indoorseg}.
Significantly, our DR-Point demonstrates a notable improvement in comparison to the Transformer trained from scratch. 
It achieves a performance gain of 2.9$\%$ in mean accuracy (mAcc) and 3.7$\%$ in mean intersection over union (mIoU).
This result serves as evidence that our DR-Point effectively enhances the Transformer's capabilities in addressing such demanding downstream tasks. 
Significantly, our DR-Point demonstrates superior performance compared to other self-supervised baselines. It achieves the highest performance by improving the mAcc and mIoU by 0.8$\%$ and 0.26$\%$ respectively, surpassing the second-best outcome achieved by Point-MAE.
In comparison to approaches that rely on scene geometric features and colors, as exemplified by the top four methods presented in Table~\ref{tab:indoorseg}, our DR-Point exhibits comparable or even better performance.

\noindent\subsection{Indoor 3D Object Detection}
Moreover, we proceeded with the evaluation of our DR-Point approach to the task of 3D object detection, which requires robust methods for understanding large-scale scenes. To achieve this, we experimented using the widely adopted real-world dataset, ScanNet V2. The results, presented in Table~\ref{tab:indoordet}, are measured in terms of $\textit{AP}_{25}$ and $\textit{AP}_{50}$. 
Through a comparison of the performance between the methods trained from scratch and those employing pre-training techniques, it becomes evident that our approach attains superior scores in terms of $\textit{AP}_{25}$ and $\textit{AP}_{50}$.

\noindent\subsection{Point Cloud Completion} 
Since previous self-supervised learning methods have mainly focused solely on the discriminant capabilities of the representation learned by the network and evaluated it by performing transfer learning to classification applications, generative capabilities of the model have been rarely studied~\citep{zhao2021relationship,zhang2022point,zhu2023csdn,yan2021consistent}. 
In this study, we verify the DR-Point's ability to perform transfer learning for point cloud completion. 
We evaluate the DR-Point on four datasets, namely PCN~\citep{yuan2018pcn}, MVP~\citep{pan2021variational}, ShapeNet55~\citep{yu2021pointr}, and ShapeNet34~\citep{yu2021pointr}, which are designed to assess the performance of point cloud completion. 
PCN is a widely used dataset with 8 categories, while MVP is presented with more classes and viewpoints. 
ShapeNet55 utilizes all categories of ShapeNet, and ShapeNet34 is usually performed to test generalization capabilities. 
As shown in Fig.~\ref{fig:PCN_vis_results},~\ref{fig:MVP_vis_results},~\ref{fig:ShapeNet55_vis_results},and~\ref{fig:Unseen21_vis_results}, DR-Point performed well in completing all partial point clouds from the four datasets, outperforming nearly all other supervised methods, such as PCN \citep{yuan2018pcn}, GRNet \citep{xie2020grnet}, TopNet \citep{tchapmi2019topnet}, and even the state-of-the-art PoinTr \citep{yu2021pointr} and SnowflakeNet \citep{xiang2021snowflakenet}. 
Table \ref{table:PCN_MVP} and \ref{table:completion-ShapeNet} shows the quantitative results, where DR-Point achieved the highest F-score@1$\%$ and lowest CD-${\ell}_1$ and CD-${\ell}_2$ in all datasets, indicating that our DR-Point performed excellently in completing point cloud data under various classes, viewpoints, and defect levels, as well as possessing strong generalization capabilities for unseen objects.

\section{Ablation Study and Analysis}

\noindent\subsection{Ablation studies on the stability of the training and the effectiveness of each individual loss}
To demonstrate the stability of the training and the effectiveness of each individual loss, we conducted additional ablation studies.
We set equal weights for the three contrastive losses to ensure a balanced contribution. As shown in Table~\ref{r1}, Models A, B, and C exhibit a decrease in performance on ModelNet40 and ScanObjectNN-BG when varying the weights of the contrastive losses. If the weights of the contrastive losses are too small, the alignment of the tri-modal representation becomes ineffective. Conversely, if the weights are too large, the model overly prioritizes the contrastive losses, neglecting the reconstruction losses.
The absence of MoCo loss (Model D) results in reduced reconstruction accuracy of the TTA branch. Similarly, removing the rendering loss (Model F) leads to a decrease in performance due to compromised reconstruction accuracy in the PTA branch.
Conversely, increasing the MoCo loss (Model E) will only result in a slight decrease in performance on downstream tasks.
Thus, we selected the optimal combination of losses as reported in the paper, and these ablation studies will be included in the revised version.

\noindent\subsection{Insight of Tri-modal Learning Objective}
DR-Point aims at pre-training the backbone by utilizing a joint learning objective. By addressing tri-modal correspondence in a unified manner can significantly enhance overall performance.
Specifically, our analysis of Table~\ref{ablation} reveals that the evaluation conducted with a tri-modal learning objective exhibits superior performance in terms of classification accuracy on ModelNet40 and ScanObjectNN (OBJ-BG), surpassing the evaluations conducted with two or one intra-modal learning objectives.
We believe that the tri-modal learning objective enhances the understanding of semantic parts by embedding the features from three modalities close to each other.

\noindent\subsection{Number of RGB and Depth Images}
We conducted an investigation to assess the impact of varying the number of rendered RGB and depth images on the performance of the two image branches. To achieve this, we selected rendered RGB and depth images captured from different random directions.
When multiple rendered RGB and depth images are available, we compute the mean of all projected features to perform tri-modal pre-training. 
The classification results for ModelNet40 are displayed in Table~\ref{ablation-images}. DR-Point showcases the ability to capture tri-modal correspondence and achieve exceptional linear classification results, even with just a single rendered RGB and depth image.
It is apparent that utilizing more than two rendered RGB and depth images can potentially introduce redundancy in the information extracted from these modalities. Consequently, this redundancy may lead to a decline in accuracy.

\noindent\subsection{Number of Depth Images}
We performed additional ablation experiments to investigate the impact of varying the number of rendered depth images.
We performed four experiments using 8, 16, 24, and 32 depth images for downstream classification, shown in Table~\ref{ablation3}. 
The purpose of this ablation study is to verify whether increasing the number of rendered depth images enhances the accuracy of reconstruction in the point-level auto-encoder. 
Additionally, besides improving reconstruction, the Transformer encoder can also benefit from enhanced learning capabilities.

\subsection{Visualization results}
In order to further gain insight into the effectiveness of DR-Point, the learned features are visualized through t-SNE~\citep{van2008visualizing}. 
Fig. \ref{fig:tsne} (\textbf{Left}) and \ref{fig:tsne} (\textbf{Right}) give the visualization of features fine-tuned on ModelNet40 and ScanObjectNN, where features form multiple clusters are well separate from each other, demonstrating the effectiveness of DR-Point.

\section{Conclusion}
We propose DR-Point, a tri-modal pre-training framework that aims to align multiple modalities, including RGB images, depth images, and point clouds, within a unified feature space. We leverage the differentiable rendering to enhance the accuracy of reconstructed point clouds and provide depth images. Experimental results demonstrate that DR-Point effectively enhances the representations of 3D backbones. 
DR-Point outperforms existing techniques on 7 point cloud processing tasks. 
Additionally, our qualitative evaluation reveals the promising potential of DR-Point for cross-modal retrieval applications.


\bibliographystyle{unsrtnat}
\bibliography{ref}  






\end{document}